\documentclass[twocolumn, aps,prd,preprintnumbers,superscriptaddress,showpacs,nofootinbib]{revtex4-1}
\usepackage{amsmath,amsthm,amssymb,mathptmx}
\usepackage[dvipdfmx]{graphicx}
\usepackage{color}
\input{colordvi.tex}
\usepackage{float}
\usepackage{booktabs}
\usepackage{bm}

\allowdisplaybreaks[1]

\newcommand{\fnleq}{f_{\rm NL}^{\rm equil}}
\newcommand{\gnleq}{g_{\rm NL}^{\rm equil}}

\newcommand{\gnl}{g_{\rm NL}}

\begin{document}
%%%%%%%%%%%%%%%%%%%%%%%%%%%%%%%%%%%%%%%%%%%%%%%%%%%%%%%%%%%%%%%%%%%%%%%%%%%%%%%%
%%%%%%%%%%%%%%%%%%%%%%%%%%%%%%%%%%%%%%%%%%%%%%%%%%%%%%%%%%%%%%%%%%%%%%%%%%%%%%%%
\title{Constraining equilateral-type primordial non-Gaussianities from  imaging surveys} 
%%%%%%%%%%%%%%%%%%%%%%%%%%%%%%%%%%%%%%%%%%%%%%%%%%%%%%%%%%%%%%%%%%%%%%%%%%%%%%%%
%%%%%%%%%%%%%%%%%%%%%%%%%%%%%%%%%%%%%%%%%%%%%%%%%%%%%%%%%%%%%%%%%%%%%%%%%%%%%%%%

\author{Ichihiko~Hashimoto}
\affiliation{Yukawa Institute for Theoretical Physics, Kyoto University, Kyoto 606-8502, Japan}

\author{Shuntaro~Mizuno}
\affiliation{Waseda Institute for Advanced Study, Waseda University, 1-6-1 Nishi-Waseda, Shinjuku, 
	Tokyo 169-8050, Japan}

\author{Shuichiro~Yokoyama}
\affiliation{Department of Physics,
Rikkyo University, 3-34-1 Nishi-Ikebukuro, Toshima, Tokyo 171-8501, Japan}

\date{\today}

%%%%%%%%%%%%%%%%%%%%%%%%%%%%%%%%%%%%%%%%%%%%%%%%%%%%%%%%%%%%%%%%%%%%%%%%%%%%%%%%
%%%%%%%%%%%%%%%%%%%%%%%%%%%%%%%%%%%%%%%%%%%%%%%%%%%%%%%%%%%%%%%%%%%%%%%%%%%%%%%%
\begin{abstract}

We investigate expected constraints on equilateral-type primordial non-Gaussianities from future/ongoing imaging surveys,
making use of the fact that they enhance the halo/galaxy bispectrum on large scales. 
As model parameters to be constrained, in addition to  $\fnleq$, which is related to the primordial bispectrum,
we consider  $\gnl^{(\partial \sigma)^4}$, which is related to the primordial trispectrum that appeared in the effective field theory of inflation.
After calculating the angular bispectra of the halo/galaxy clustering and weak gravitational lensing based on the integrated perturbation theory, we perform Fisher matrix analysis for three representative surveys. 
We find that among the three surveys, the tightest constraints come from Large Synoptic Survey Telescope ; its expected $1\sigma$ errors on $\fnleq$ and $\gnl^{(\partial \sigma)^4}$ are respectively given by $7.0 \times 10^2$ and $4.9 \times 10^7$.
Although this constraint is somewhat looser than the one from the current cosmic microwave background observation, 
since we obtain it independently, we can use this constraint as a cross check. We also evaluate the uncertainty with our results caused by using
several approximations and discuss the possibility to obtain tighter constraint on $\fnleq$ and $\gnl^{(\partial \sigma)^4}$.
 
\end{abstract}
%%%%%%%%%%%%%%%%%%%%%%%%%%%%%%%%%%%%%%%%%%%%%%%%%%%%%%%%%%%%%%%%%%%%%%%%%%%%%%%%
%%%%%%%%%%%%%%%%%%%%%%%%%%%%%%%%%%%%%%%%%%%%%%%%%%%%%%%%%%%%%%%%%%%%%%%%%%%%%%%%

\preprint{RUP-16-14}
\preprint{YITP-16-65}

\maketitle

%\tableofcontents

%%%%%%%%%%%%%%%%%%%%%%%%%%%%%%%%%%%%%%%%%%%%%%%%%%%%%%%%%%%%%%%%%%%%%%%%%%%%%%%%
%%%%%%%%%%%%%%%%%%%%%%%%%%%%%%%%%%%%%%%%%%%%%%%%%%%%%%%%%%%%%%%%%%%%%%%%%%%%%%%%
\section{Introduction}
\label{sec:intro}

It has been widely known that the primordial non-Gaussianity is a powerful tool to understand the non-linearity and the interaction structure of inflationary era
(for a review, see \cite{Bartolo:2004if}).
There are a large number of theoretical and observational studies for the primordial non-Gaussianity.
Currently, the most stringent observational constraints on the primordial non-Gaussianity have been obtained from the analysis of the higher-order spectra of the cosmic microwave background (CMB)
anisotropies performed by Planck Collaboration \cite{2015arXiv150201592P}, and they show no evidence of the primordial non-Gaussianity, which is consistent with the standard single slow-roll inflation model.
However, from the theoretical point of view 
the constraints are still somewhat weak, and 
it would be interesting and important to investigate the observational constraint  independently from the CMB observations.

Alternative information  which can be expected to probe the primordial non-Gaussian feature is obtained through large-scale structure observations.
As the effect of the primordial non-Gaussianity, it has been known that
the power spectrum of the biased tracers, such as haloes/galaxies, could be enhanced on large scales
compared with the purely Gaussian case, which is called as a ``scale-dependent bias" feature (e.g., \cite{Dalal:2007cu,Slosar:2008hx,Matarrese:2008nc}).
However, such an enhancement could be realized for the so-called ``local-type" primordial non-Gaussianity,
which can be produced in some multi-field inflation models \cite{Wands:2010af,Suyama:2010uj,Byrnes:2010em}.
On the other hand, there is another interesting class of primordial non-Gaussianity,
so-called ``equilateral-type" one, which would be large in
non-canonical field driven inflation models  \cite{Koyama:2010xj,Chen:2010xka}. 
It is shown that this type of primordial non-Gaussianity gives no distinct scale-dependence in the power spectrum,
and hence it had seemed to be difficult to obtain a significant constraint on the equilateral-type non-Gaussianity from the large-scale structure observations.

Regardless of this, recently, there have been several works which discuss the possibility of probing the equilateral-type non-Gaussianity
from the large-scale structure observations  through the analysis of the higher-order spectra, e.g., bispectrum of the biased tracers.
Actually, in the presence of the equilateral-type primordial bispectrum, it has been shown that
the amplitude of the halo/galaxy bispectrum is enhanced on large scales \cite{Sefusatti:2007ih,Sefusatti:2009qh,Yokoyama:2013mta},
and making use of this fact,
 the future galaxy surveys could be expected to give a strong constraint on $\fnleq$,
which is comparable to ones obtained by CMB observations  \cite{Sefusatti:2007ih}.

In general, the equilateral-type primordial non-Gaussianity is characterized not only by $\fnleq$,
but also by  $\gnleq$, which is related to the primordial trispectrum.  
Although the shapes of primordial trispectra of this class are very complicated,
since they strongly depend on the theoretical models \cite{Mizuno:2010by,Izumi:2011di}, obtaining the observational constraint on the primordial trispectrum
is still important to clarify the interaction structure of the inflation model. 
In this respect, among the shapes of primordial trispectra, only three shapes of the primordial trispectra
appeared in the effective field theory of inflation \cite{Senatore:2010jy,Senatore:2010wk} and 
whose amplitudes are characterized by $g_{\rm NL} ^{\dot{\sigma}^4}$, $g_{\rm NL} ^{\dot{\sigma}^2 (\partial \sigma)^2}$, and
$g_{\rm NL} ^{(\partial \sigma)^4}$ have been constrained systematically by the CMB observations  \cite{Smith:2015uia}
 (we will show the detailed forms of these trispectra later
in Sec.~\ref{sec:powerandbi}). For these equilateral-type primordial trispectra, in the previous paper  \cite{2015PhRvD..91l3521M},
two of us have investigated their impacts on the halo/galaxy bispectrum and showed that 
$\gnl^{\dot{\sigma}^2 (\partial \sigma)^2}$ and $\gnl^{(\partial \sigma)^4}$ could be constrained from the halo/galaxy bispectrum observations,
while we can focus only on  $\gnl^{(\partial \sigma)^4}$ as these shapes are very close to each other
and only one of the two can be used as the basis of the optimal analysis.

In this paper, we focus on the future/ongoing imaging surveys and investigate the expected constraints on the equilateral-type primordial non-Gaussianity
from the analysis of the halo/galaxy bispectrum and estimate $1\sigma$ errors on  $\fnleq$ and $\gnl^{(\partial \sigma)^4}$.
In our analysis, as in the previous work \cite{Hashimoto:2015tnv} where two of us were involved, 
we also include the cross correlation between halo/galaxy density field and the weak gravitational lensing which directly traces
the matter density field. 

The paper is organized as follows. In Sec.~\ref{sec:powerandbi}, we present formulas for the three dimensional auto-/cross-bispectra of halo/galaxy
and matter distribution in the presence of the equilateral-type primordial non-Gaussianity. 
We also derive angular bispectra
which are observables of the photometric/imaging galaxy surveys. 
In Sec.~\ref{sec:fifi}, based on the Fisher matrix formalism,
we quantitatively estimate the impact of the bispectra and weak-gravitational lensing effect
on the detection of the equilateral-type primordial non-Gaussianity. Finally, Sec.~\ref{sec:sum} is devoted to summary and discussion.
Throughout this paper, unless specifically mentioned, we adopt the best fit cosmological parameters taken from Plank \cite{2015arXiv150201589P}.

%%%%%%%%%%%%%%%%%%%%%%%%%%%%%%%%%%%%%%%%%%%%%%%%%%%%%%%%%%%%%%%%%%%%%%%%%%%%%%%%
%%%%%%%%%%%%%%%%%%%%%%%%%%%%%%%%%%%%%%%%%%%%%%%%%%%%%%%%%%%%%%%%%%%%%%%%%%%%%%%%
%%%%%%%%%%%%%%%%%%%%%%%%%%%%%%%%%%%%%%%%%%%%%%%%%%%%%%%%%%%%%%%%%%%%%%%%%%%
%%%%%%%%%%%%%%%%%%%%%%%%%%%%%%%%%%%%%%%%%%%%%%%%%%%%%%%%%%%%%%%%%%%%%%%%%%%
\section{Halo/Galaxy and weak lensing bispectra and with equilateral-type primordial non-Gaussianities\label{sec:powerandbi}}
%%%%%%%%%%%%%%%%%%%%%%%%%%%%%%%%%%%%%%%%%%%%%%%%%%%%%%%%%%%%%%%%%%%%%%%%%%%
%%%%%%%%%%%%%%%%%%%%%%%%%%%%%%%%%%%%%%%%%%%%%%%%%%%%%%%%%%%%%%%%%%%%%%%%%%%

%%%%%%%%%%%%%%%%%%%%%%%%%%%%%%%%%%%%%%%%%%%%%%%%%%%%%%%%%%%%%%%%%%%%%%%%%%%%%%%%
\subsection{Equilateral-type primordial non-Gaussianities\label{sec:equil}}
%%%%%%%%%%%%%%%%%%%%%%%%%%%%%%%%%%%%%%%%%%%%%%%%%%%%%%%%%%%%%%%%%%%%%%%%%%%%%%%%

Here, we present the concrete form of the statistical quantities of primordial curvature perturbations as well as
those for the linear density for the situations we are interested in. 

We begin by defining the power spectrum of the primordial curvature perturbation,  $P_{\Phi}$,
\begin{equation}
\langle \Phi ( {\bf k}) \Phi  ( {\bf k}') \rangle = (2 \pi)^3 \delta^{(3)} _{\rm D}( {\bf k} + {\bf k}' ) P_{\Phi} (k)\,,
\end{equation}
where $ {\bf k}$ and $ \delta^{(3)} _{\rm D}$ are three-dimensional wave vector and the three-dimensional Dirac's delta function, respectively.
The bracket $\langle \rangle$ means the ensemble average. If the curvature perturbation is generated by a single-field slow-roll inflation, it can be shown that
these perturbations almost obey Gaussian statistics, and thus the statistical properties are completely characterized by the power spectrum 
\cite{Maldacena:2002vr}.

On the other hand, if we consider inflation models which are no longer the single-field slow-roll ones, like 
models with multiple fields, non slow-roll background dynamics and  non-canonical kinetic terms, 
non-Gaussianities of primordial perturbations are generated.
In such cases, the non-Gaussian nature of the primordial perturbations is encoded 
in the higher-order spectra of primordial curvature perturbations such as the bispectrum, $B_{\Phi}$, and trispectrum, $T_{\Phi}$:
\begin{eqnarray}
&&\langle \Phi ( {\bf k}_1) \Phi  ( {\bf k}_2)  \Phi  ( {\bf k}_3)  \rangle_{\rm c} \nonumber\\
&&\hspace{1cm}=  (2 \pi)^3\delta^{(3)} _{\rm D}( {\bf k}_1 +  {\bf k}_2 +  {\bf k}_3 ) B_{\Phi} ( {\bf k}_1, \;{\bf k}_2, \;{\bf k}_3 )\,,\\
&&\langle \Phi ( {\bf k}_1) \Phi  ( {\bf k}_2)  \Phi  ( {\bf k}_3)   \Phi  ( {\bf k}_4)  \rangle_{\rm c} \nonumber\\
&&\hspace{1cm}=  (2 \pi)^3 \delta^{(3)} _{\rm D}( {\bf k}_1 +  {\bf k}_2 +  {\bf k}_3  +  {\bf k} _4) T_{\Phi} ( {\bf k}_1, \;{\bf k}_2, \;{\bf k}_3,  \;{\bf k}_4 )\,,
\end{eqnarray}
 where the subscript, ${\rm c}$, means that we consider only the connected part of the correlation functions.

Among several types of primordial non-Gaussianities known so far, we concentrate in this paper on the equilateral-type one
that could be produced by inflation models with non-canonical kinetic terms where the non-linear interactions 
become important on subhorizon scales.
 (For concrete inflation models
which produce equilateral-type primordial non-Gaussianity, see reviews, e.g. \cite{Koyama:2010xj,Chen:2010xka}.)

In this class of inflation models, it was shown that the bispectrum of the primordial curvature perturbation
is well approximated by the following separable form in most cases \cite{Creminelli:2005hu}: 
\begin{eqnarray}
&&B_{\Phi} ^{\rm equil}   ({\bf k}_1,  {\bf k}_2,{\bf k}_3)\nonumber\\
&&\hspace{0.5cm} =6 f_{\rm NL} ^{\rm equil}  
[ -(P_\Phi (k_1) P_\Phi (k_2) + 2\; {\rm perms.}   )\nonumber\\
&&\hspace{1cm}-2 P_\Phi (k_1) ^{2/3} P_\Phi (k_2) ^{2/3}  P_\Phi (k_3) ^{2/3}\nonumber\\
&&\hspace{1cm}
+( P_\Phi (k_1) ^{1/3}   P_\Phi (k_2) ^{2/3} P_\Phi (k_3)     + 5\; {\rm perms.}     ) ]\,,
\label{shape_equilateralbispectrum}
\end{eqnarray} 
which is called the equilateral-type primordial bispectrum. Here $ f_{\rm NL} ^{\rm equil}  $
is the nonlinearity parameter which characterizes the amplitude of the bispectrum.

On the other hand, the form of primordial trispectrum generated by this class of inflation models strongly depends on the inflation models
and it is very complicated in general. Recently, however, for relatively simple primordial  trispectra in this class, the constraints on their amplitudes
have been obtained by CMB observations \cite{Smith:2015uia}. The concrete forms of the primordial trispectra investigated
in the work are given by
\begin{eqnarray}
&&T_\Phi ^{\dot{\sigma}^4}  ({\bf k}_1,  {\bf k}_2,{\bf k}_3,{\bf k}_4)\nonumber \\
&&\hspace{0.5cm} = \frac{221184}{25}\; g_{\rm NL} ^{\dot{\sigma}^4}\;A_\Phi^3\;S^{\dot{\sigma}^4}
 ({\bf k}_1,  {\bf k}_2,{\bf k}_3,{\bf k}_4)\,,\label{tris_c1}\\
&&T_\Phi ^{\dot{\sigma}^2 (\partial \sigma)^2}  ({\bf k}_1,  {\bf k}_2,{\bf k}_3,{\bf k}_4)\nonumber\\
&&\hspace{0.5cm} = -\frac{27648}{325}\; g_{\rm NL} ^{\dot{\sigma}^2 (\partial \sigma)^2}\;  A_\Phi^3\;
S^{\dot{\sigma}^2 (\partial \sigma)^2} ({\bf k}_1,  {\bf k}_2,{\bf k}_3,{\bf k}_4) \,,\label{tris_c2}\\
&&T_\Phi ^{(\partial \sigma)^4}  ({\bf k}_1,  {\bf k}_2,{\bf k}_3,{\bf k}_4)\nonumber\\ 
&& \hspace{0.5cm} = \frac{16588}{2575} \;g_{\rm NL} ^{(\partial \sigma)^4} \; A_\Phi^3\; S^{(\partial \sigma)^4}
 ({\bf k}_1,  {\bf k}_2,{\bf k}_3,{\bf k}_4) \,,\label{tris_c3}
\end{eqnarray} 
with
\begin{eqnarray}
&&S^{\dot{\sigma}^4}  ({\bf k}_1,  {\bf k}_2,{\bf k}_3,{\bf k}_4) = \frac{1}{\left(\sum_{i=1} ^4 k_i \right)^5 \Pi_{i=1} ^4 k_i}\,,\label{shape_c1}\\
&&S^{\dot{\sigma}^2 (\partial \sigma)^2}  ({\bf k}_1,  {\bf k}_2,{\bf k}_3,{\bf k}_4)\nonumber\\
&&  \hspace{0.5cm}= \frac{k_1^2 k_2^2 ( {\bf k}_3 \cdot  {\bf k}_4)}{\left(\sum_{i=1} ^4 k_i \right)^3 \Pi_{i=1} ^4 k_i ^3}
\left( 1 + 3 \frac{k_3 + k_4}{   \sum_{i=1} ^4 k_i    } +12 \frac{k_3  k_4}{\left(\sum_{i=1} ^4 k_i \right)^2 }\right)\nonumber\\
&&  \hspace{1cm} +5\;\; {\rm perms.}\,,\label{shape_c2}\\
&&S^{(\partial \sigma)^4}  ({\bf k}_1,  {\bf k}_2,{\bf k}_3,{\bf k}_4)\nonumber\\ 
&& = \frac{  ({\bf k}_1 \cdot  {\bf k}_2) ({\bf k}_3 \cdot  {\bf k}_4) +  ({\bf k}_1 \cdot  {\bf k}_3) ({\bf k}_2 \cdot  {\bf k}_4) +
 ({\bf k}_1 \cdot  {\bf k}_4) ({\bf k}_2 \cdot  {\bf k}_3) }{ \sum_{i=1} ^4 k_i  \Pi_{i=1} ^4 k_i ^3 }\nonumber\\
&& \times \left(1 + \frac{ \sum_{i<j} k_i k_j}{\left(\sum_{i=1} ^4 k_i \right)^2} + 
3 \frac{  \Pi_{i=1} ^4 k_i }{\left(\sum_{i=1} ^4 k_i \right)^3}   \sum_{i=1} ^4 \frac{1}{k_i} + 
12 \frac{  \Pi_{i=1} ^4 k_i }{   \left(\sum_{i=1} ^4 k_i \right)^4 }\right)\,.\nonumber\\
\label{shape_c3}
\end{eqnarray} 
Here, the parameters $g_{\rm NL} ^{\dot{\sigma}^4}$, $g_{\rm NL} ^{\dot{\sigma}^2 (\partial \sigma)^2}$ and $g_{\rm NL} ^{(\partial \sigma)^4}$
describe the strength of the primordial non-Gaussianity, $ A_\Phi$ is the amplitude of the primordial power spectrum, defined by 
$A_\Phi = k^3 P_\Phi$ and the normalizations are the ones adopted in Ref.~\cite{Smith:2015uia}.
It is worth mentioning that these primordial trispectra are not only relatively simple, but also have natural theoretical origin in the sense that
they are shown to be generated by   the effective field theory of inflation \cite{Senatore:2010jy,Senatore:2010wk} 
as well as $k$ inflation  \cite{Huang:2006eha,Arroja:2009pd,Chen:2009bc}.

Although these three primordial trispectra are equally important in the context of the effective field theory of inflation, it was shown that 
because the primordial trispectra $T_\Phi ^{(\partial \sigma)^4}$ and $T_\Phi ^{\dot{\sigma}^2 (\partial \sigma)^2}$ have similar shape dependence, 
only two of them can be used as the basis of  the optimal analysis of the CMB trispectrum  \cite{Smith:2015uia}.
Furthermore, in \cite{2015PhRvD..91l3521M}, two of us showed that  $ g_{\rm NL} ^{\dot{\sigma}^4}$ 
can not be constrained from the observations of Halo/galaxy bispectrum as this contribution
never dominates the one from the gravitational nonlinearity on large scales. From these reasons, we concentrate on
the primordial trispectrum $T_\Phi ^{(\partial \sigma)^4} $ that is characterized by Eqs.~(\ref{tris_c3}) and (\ref{shape_c3})
and, for brevity we call this primordial trispectrum as equilateral-type trispectrum through this paper.

Based on the primordial curvature perturbation satisfying the statistical properties discussed above, 
the linear density field $\delta_{\rm L}$is obtained through 
\begin{eqnarray}
&&\delta_{\rm L} ({\bf k},\;z) = M (k,\;z) \Phi  ({\bf k},\;z) \label{rel_dell_Phi};\,\\
&&M  (k,\;z)  = \frac23 \frac{D(z)}{D(z_*) (1+z_*)} \frac{k^2 T(k)}{H_0 ^2 \Omega_{\rm m0}}\,,
\label{def_m}
\end{eqnarray}
where we relate the linear density field to the primordial curvature perturbations by a function $M (k,\;z)$, which is given by the transfer function $T(k)$ and the linear growth factor $D(z)$. The exact formula of linear growth factor is determined from linear theory, and the transfer function are computed from CAMB \cite{2000ApJ...538..473L}. $H_0$, and $\Omega_{\rm m0}$ are the Hubble parameter at present epoch and the matter density parameter, respectively and $z_*$ denotes an arbitrary redshift at the matter-dominated era.
Then, the power-, bi-, and tri-spectra of the linear density field 
are defined by
\begin{eqnarray}
&&\langle \delta_{\rm L} (  {\bf k}_1)  \delta_{\rm L} ( {\bf k}_2) \rangle = (2 \pi)^3 \delta_{\rm D}^{(3)}( {\bf k}_1 +  {\bf k}_2 ) P_{\rm L} (k)\,,\\ 
&&\langle \delta_{\rm L} ( {\bf k}_1)  \delta_{\rm L} ( {\bf k}_2)  \delta_{\rm L} ( {\bf k}_3) \rangle\nonumber\\
&&\hspace{0.5 cm}= (2 \pi)^3 \delta_{\rm D}^{(3)}( {\bf k}_1 + {\bf k}_2 + {\bf k}_3 ) 
B_{\rm L} ( {\bf k}_1, {\bf k}_2, {\bf k}_3)\,\\
&&\langle \delta_{\rm L} ( {\bf k}_1)  \delta_{\rm L} ( {\bf k}_2)  \delta_{\rm L} ( {\bf k}_3)  \delta_{\rm L} ( {\bf k}_4) \rangle \nonumber\\
&&\hspace{0.5 cm}= 
(2 \pi)^3 \delta_{\rm D}^{(3)}( {\bf k}_1 + {\bf k}_2 +  {\bf k}_3  +  {\bf k}_4 ) 
T_{\rm L} ( {\bf k}_1, {\bf k}_2, {\bf k}_3 , {\bf k}_4)\,.
\end{eqnarray}

From Eqs.~(\ref{rel_dell_Phi}) and (\ref{def_m}), we can relate these spectra to those of the primordial curvature perturbations as
\begin{eqnarray}
&&P_{\rm L} (k)=  M (k)^2P_{\Phi} (k) \,,\\ 
&&B_{\rm L} ( {\bf k}_1, {\bf k}_2, {\bf k}_3) =   M (k_1) M (k_2)  M (k_3) B_{\Phi}  ( {\bf k}_1, {\bf k}_2, {\bf k}_3) \,,\label{linear_bispec}\\
&&
T_{\rm L} ( {\bf k}_1, {\bf k}_2, {\bf k}_3 , {\bf k}_4)\nonumber\\
&&\hspace{0.5 cm} =  M (k_1) M (k_2)  M (k_3) M (k_4)T_{\Phi} ( {\bf k}_1, {\bf k}_2, {\bf k}_3 , {\bf k}_4)\,.\label{linear_triispec}
\end{eqnarray}

Equipped with the linear density field presented in this subsection, we will derive in the next subsection the observables of large-scale structure
which are probed with future imaging surveys.

%%%%%%%%%%%%%%%%%%%%%%%%%%%%%%%%%%%%%%%%%%%%%%%%%%%%%%%%%%%%%%%%%%%%%%%%%%%
\subsection{Halo/galaxy and weak lensing bispectra based on the Integrated Perturbation Theory (iPT) \label{sec:ipt}}
%%%%%%%%%%%%%%%%%%%%%%%%%%%%%%%%%%%%%%%%%%%%%%%%%%%%%%%%%%%%%%%%%%%%%%%%%%%

In addition to the halo/galaxy bispectrum with primordial equilateral-type non-Gaussianities discussed in  \cite{2015PhRvD..91l3521M},
here we consider the cross bispectra between the halo/galaxy density field and the weak gravitational lensing.
Let us first define the three-dimensional bispectra $\mathcal{B}_{\rm XYZ}$ 
of the observables: 
%%%%%%%%%%%%%%%%%%%%%%%%%%%%%%%%%%%%%%%%%%%%%%%%%%%%%%%%%%%%%%%%%%%%%%%%%%%
\begin{align}
&\frac{1}{3}\left\{\langle \delta_{\rm X}(\bm{k}_1)\delta_{\rm Y}(\bm{k}_2)\delta_{\rm Z}(\bm{k}_3)\rangle+\mbox{2 perms}\,(\bm{k}_1\leftrightarrow \bm{k}_2\leftrightarrow\bm{k}_3)\right\}
\nonumber\\
&\qquad\qquad=(2\pi)^3\delta_{\rm D}^{(3)}(\bm{k}_1+\bm{k}_2+\bm{k}_3)\,\mathcal{B}_{\rm XYZ}(k_1,k_2,k_3)
,\label{bis3}
\end{align}
%%%%%%%%%%%%%%%%%%%%%%%%%%%%%%%%%%%%%%%%%%%%%%%%%%%%%%%%%%%%%%%%%%%%%%%%%%%
where $\delta_{\rm X,Y,Z}$ is the three-dimensional density field,
and $X, Y, Z = {\rm h}, {\rm m}$ respectively represents the halo/galaxy density field and the matter fluctuation. 
Here, we employ the integrated Perturbation Theory (iPT) \cite{2011PhRvD..83h3518M,2012PhRvD..86f3518M} to obtain the analytic expression for the bispectra $\mathcal{B}_{\rm XYZ}$ in terms of the primordial non-Gaussianities. 
In iPT,
the statistical quantities such as power spectra and bispectra 
of the halo/galaxy density field and matter fluctuations
are perturbatively constructed with the linear polyspectra 
and  multi-point propagators which can be defined as~ \cite{2011PhRvD..83h3518M,2012PhRvD..86f3518M}

%%%%%%%%%%%%%%%%%%%%%%%%%%%%%%%%%%%%%%%%%%%%%%%%%%%%%%%%%%%%%%%%%%%%%%%%%%%
\begin{align}
&\left\langle \frac{\delta^n\delta_{\rm X}\left(\bm{k}\right)}{\delta\delta_{\rm L}\left(\bm{k}_1\right)\delta\delta_{\rm L}\left(\bm{k}_2\right)\cdots\delta\delta_{\rm L}\left(\bm{k}_n\right)}\right\rangle
\nonumber\\
&\quad\quad=
\left(2\pi\right)^{3-3n}\delta_{\rm D}^{(3)}\left(\bm{k}-\bm{k}_{12\cdots n}\right)\Gamma_{\rm X}^{\left(n\right)}\left(\bm{k}_1,\bm{k}_2,\cdots,\bm{k}_n\right)
\label{gamma}\, ,
\end{align}
%%%%%%%%%%%%%%%%%%%%%%%%%%%%%%%%%%%%%%%%%%%%%%%%%%%%%%%%%%%%%%%%%%%%%%%%%%%
where ${\bm k}_{12\cdots n} = {\bm k}_1 + {\bm k}_2 + \cdots + {\bm k}_n$.
With these propagators, the bispectra with the primordial non-Gaussianities are  expressed as
%%%%%%%%%%%%%%%%%%%%%%%%%%%%%%%%%%%%%%%%%%%%%%%%%%%%%%%%%%%%%%%%%%%%%%%%%%%
\begin{align}
&\mathcal{B}_{\rm XYZ}(k_1,k_2,k_3)=\mathcal{B}^{\rm grav}_{\rm XYZ}(k_1,k_2,k_3)+\mathcal{B}^{\rm bis}_{\rm XYZ}(k_1,k_2,k_3) \nonumber\\
& \quad\quad  +\mathcal{B}^{\rm tris}_{\rm XYZ}(k_1,k_2,k_3).\label{bis3D}
\end{align}
%%%%%%%%%%%%%%%%%%%%%%%%%%%%%%%%%%%%%%%%%%%%%%%%%%%%%%%%%%%%%%%%%%%%%%%%%%%
Here, the quantities $\mathcal{B}^{\rm grav}_{\rm XYZ}$, $\mathcal{B}^{\rm bis}_{\rm XYZ}$ and $\mathcal{B}^{\rm tris}_{\rm XYZ}$
are respectively corresponding to the contributions from the non-linear gravitational evolution, the equilateral-type primordial bispectrum characterized by $\fnleq$,
and the equilateral-type primordial trispectrum characterized by 
$g_{\rm NL} ^{(\partial \sigma)^4}$, as shown in the previous subsection.

The explicit expression for each contribution is given by
%%%%%%%%%%%%%%%%%%%%%%%%%%%%%%%%%%%%%%%%%%%%%%%%%%%%%%%%%%%%%%%%%%%%%%%%%%%
\begin{widetext}
\begin{align}
\mathcal{B}^{\rm grav}_{XYZ}\left(\bm{k}_1,\bm{k}_2,\bm{k}_3\right)=&\frac{1}{3}\Bigl[\Bigl\{\Gamma^{\left(1\right)}_{\rm X}\left(\bm{k}_1\right)\Gamma^{\left(1\right)}_{\rm Y}\left(\bm{k}_2\right)\Gamma^{\left(2\right)}_{\rm Z}\left(-\bm{k}_1,-\bm{k}_2\right)P_{\rm L}\left(k_1\right)P_{\rm L}\left(k_2\right)\nonumber\\
&+2{\rm perms}({\rm X\leftrightarrow Y\leftrightarrow Z})\Bigr\}+2{\rm perms}(\bm{k}_1\leftrightarrow \bm{k}_2\leftrightarrow \bm{k}_3)\Bigr],\label{treegrav}\\
\mathcal{B}^{\rm bis}_{XYZ}\left(\bm{k}_1,\bm{k}_2,\bm{k}_3\right)=&\Gamma^{\left(1\right)}_{\rm X}\left(\bm{k}_1\right)\Gamma^{\left(1\right)}_{\rm Y}\left(\bm{k}_2\right)\Gamma^{\left(1\right)}_{\rm Z}\left(\bm{k}_3\right)B_{\rm L}\left(\bm{k}_1,\bm{k}_2,\bm{k}_3\right),\label{treebis}\\
\mathcal{B}_{XYZ}^{\rm tris}\left(\bm{k}_1,\bm{k}_2,\bm{k}_3\right)=&\frac{1}{3}\Biggl[\Biggl\{\frac{1}{2}\Gamma^{\left(1\right)}_{\rm X}\left(\bm{k}_1\right)\Gamma^{\left(1\right)}_{\rm Y}\left(\bm{k}_2\right)\int\frac{d^3p}{\left(2\pi\right)^3}\Gamma^{\left(2\right)}_{\rm Z}\left(\bm{p},\bm{k}_3-\bm{p}\right)T_{\rm L}\left(\bm{k}_1,\bm{k}_2,\bm{p},\bm{k}_3-\bm{p}\right)\nonumber\\
&+2{\rm perms}({\rm X\leftrightarrow Y\leftrightarrow Z})\Biggr\}+2{\rm perms}(\bm{k}_1\leftrightarrow \bm{k}_2\leftrightarrow \bm{k}_3)\Biggr].\label{tris}
\end{align}
\end{widetext}
%%%%%%%%%%%%%%%%%%%%%%%%%%%%%%%%%%%%%%%%%%%%%%%%%%%%%%%%%%%%%%%%%%%%%%%%%%%
Here, we have  neglected the higher-order contributions, which are expected to be small on large scales for the equilateral-type primordial non-Gaussianities.
Substituting the expressions for $B_{\rm L}$ and $T_{\rm L}$ given by Eqs.~(\ref{linear_bispec}) and (\ref{linear_triispec}) 
%in the previous subsection 
into the above expressions,
in the large scale limit where the scale of interest $\sim 1/k_i $ is much larger than the typical scale of the formation of the collapsed object $\sim 1/p$, we have
%%%%%%%%%%%%%%%%%%%%%%%%%%%%%%%%%%%%%%%%%%%%%%%%%%%%%%%%%%%%%%%%%%%%%%%%%%%
\begin{widetext}  
\begin{align}
\mathcal{B}^{\rm bis}_{\rm XYZ}\left(\bm{k}_1,\bm{k}_2,\bm{k}_3\right)=&6f_{\rm NL}^{\rm equil}\Gamma^{\left(1\right)}_{\rm X}\left(\bm{k}_1\right)\Gamma^{\left(1\right)}_{\rm Y}\left(\bm{k}_2\right)\Gamma^{\left(1\right)}_{\rm Z}\left(\bm{k}_3\right)M\left(k_1\right)M\left(k_2\right)M\left(k_3\right)\,\,
\biggl[-\left(P_{\Phi}\left(k_1\right)P_\Phi\left(k_2\right)+2{\rm perms}(\bm{k}_1\leftrightarrow \bm{k}_2\leftrightarrow \bm{k}_3)\right)\nonumber\\
-&2P_{\Phi}\left(k_1\right)^{2/3}P_\Phi\left(k_2\right)^{2/3}P_\Phi\left(k_3\right)^{2/3}
+\left(P_{\Phi}\left(k_1\right)^{1/3}P_\Phi\left(k_2\right)^{2/3}P_\Phi\left(k_3\right)+5{\rm perms}(\bm{k}_1\leftrightarrow \bm{k}_2\leftrightarrow \bm{k}_3)\right)\biggr],\label{Btreebis-L}\\
\mathcal{B}^{\rm tris}_{\rm XYZ}\left(\bm{k}_1,\bm{k}_2,\bm{k}_3\right)\simeq&-\frac{4147}{6180}g_{\rm NL}^{(\partial\sigma)^4}\Biggl[\Biggl\{\Gamma^{\left(1\right)}_{\rm X}\left(\bm{k}_1\right)\Gamma^{\left(1\right)}_{\rm Y}\left(\bm{k}_2\right)M\left(k_1\right)M\left(k_2\right)P_{\Phi}\left(k_1\right)P_\Phi\left(k_2\right)\int\frac{d^3p}{\left(2\pi\right)^3}\Gamma^{\left(2\right)}_{\rm Z}\left(\bm{p},-\bm{p}\right)\frac{P_{\rm L}\left(p\right)}{p^2}\nonumber\\
\times&\left\{(\bm{k}_1\cdot\bm{k}_2)+2(\frac{\bm{p}}{p}\cdot\bm{k}_1)(\frac{\bm{p}}{p}\cdot\bm{k}_2)\right\}+2{\rm perms}({\rm X\leftrightarrow Y\leftrightarrow Z})\Biggr\}+2{\rm perms}(\bm{k}_1\leftrightarrow \bm{k}_2\leftrightarrow \bm{k}_3)\Biggr]\nonumber\\
=&-\frac{4147}{3708}g_{\rm NL}^{(\partial\sigma)^4}\Biggl[\Biggl\{\Gamma^{\left(1\right)}_{\rm X}\left(\bm{k}_1\right)\Gamma^{\left(1\right)}_{\rm Y}\left(\bm{k}_2\right)M\left(k_1\right)M\left(k_2\right)P_{\Phi}\left(k_1\right)P_\Phi\left(k_2\right)(\bm{k}_1\cdot\bm{k}_2)\int\frac{d^3p}{\left(2\pi\right)^3}\Gamma^{\left(2\right)}_{\rm Z}\left(\bm{p},-\bm{p}\right)\frac{P_{\rm L}\left(p\right)}{p^2}\nonumber\\
+&2{\rm perms}({\rm X\leftrightarrow Y\leftrightarrow Z})\Biggr\}+2{\rm perms}(\bm{k}_1\leftrightarrow \bm{k}_2\leftrightarrow \bm{k}_3)\Biggr]
.\label{1-loopgnlB}
\end{align}
\end{widetext}

As shown in Ref.~\cite{2015PhRvD..91l3521M}, 
the contributions of the equilateral-type primordial non-Gaussianities become larger on larger scales.
Hence, according to Ref.~\cite{2014PhRvD..89d3524Y}, 
by employing the large-scale limit $(k_i \to 0)$,
the multi-point propagators for halo/galaxy, which characterize the non-linear gravitational evolution and halo/galaxy bias properties, can be simply given by
%%%%%%%%%%%%%%%%%%%%%%%%%%%%%%%%%%%%%%%%%%%%%%%%%%%%%%%%%%%%%%%%%%%%%%%%%%%
\begin{align}
\Gamma^{(1)}_{\rm h}(\bm{k})&\simeq 1+c^{\rm L}_1(k),\label{gamma1}\\
\Gamma^{(2)}_{\rm h}(\bm{k}_1,\bm{k}_2)&\simeq F_2(\bm{k}_1,\bm{k}_2)+\left(1+\frac{\bm{k}_1\cdot\bm{k}_2}{k_2^2}\right)c_1^{\rm L}(\bm{k}_1)
\nonumber\\
&+\left(1+\frac{\bm{k}_1\cdot\bm{k}_2}{k_1^2}\right)c_1^{\rm L}(\bm{k}_2)+c_2^{\rm L}(\bm{k}_1,\bm{k}_2). \label{gamma2}
\end{align}
%%%%%%%%%%%%%%%%%%%%%%%%%%%%%%%%%%%%%%%%%%%%%%%%%%%%%%%%%%%%%%%%%%%%%%%%%%%
 Here  $c_n^{\rm L}$  represent the renormalized-bias function defined in Lagrangian space, which can be defined in terms of
the three-dimensional density field in Lagrangian space $\delta_{\rm X}^{\rm L}$ as
%%%%%%%%%%%%%%%%%%%%%%%%%%%%%%%%%%%%%%%%%%%%%%%%%%%%%%%%%%%%%%%%%%%%%%%%%%%%
\begin{align}
c_n^{\rm L}(\bm{k}_1,\bm{k}_2,\cdots,\bm{k}_n)&=(2\pi)^{3n}\int\frac{d^3k'}{(2\pi)^3}
\nonumber\\
&\quad\times
\left\langle \frac{\delta^n\delta_{\rm X}^{\rm L}(\bm{k}')}{\delta\delta_{\rm L}(\bm{k}_1)\delta\delta_{\rm L}(\bm{k}_2)\cdots\delta\delta_{\rm L}(\bm{k}_n)}\right\rangle.
\label{cnl}
\end{align}
%%%%%%%%%%%%%%%%%%%%%%%%%%%%%%%%%%%%%%%%%%%%%%%%%%%%%%%%%%%%%%%%%%%%%%%%%%%%
$F_2$ is the  second-order kernel of standard perturbation theory, which is given by 
%%%%%%%%%%%%%%%%%%%%%%%%%%%%%%%%%%%%%%%%%%%%%%%%%%%%%%%%%%%%%%%%%%%%%%%%%%%%
\begin{align}
F_2(\bm{k}_1,\bm{k}_2)&=\frac{10}{7}+\left(\frac{k_2}{k_1}+\frac{k_1}{k_2}\right)\frac{\bm{k}_1\cdot\bm{k}_2}{k_1k_2}+\frac{4}{7}\left(\frac{\bm{k}_1\cdot\bm{k}_2}{k_1k_2}\right)^2\,.
\end{align}

In order to obtain more concrete expressions for the renormalized bias function,
here, we adopt the halo-bias prescription proposed by \cite{2011PhRvD..83h3518M}: 

%%%%%%%%%%%%%%%%%%%%%%%%%%%%%%%%%%%%%%%%%%%%%%%%%%%%%%%%%%%%%%%%%%%%%%%%%%%
\begin{align}
&c_n^{\rm L}(\bm{k}_1,\cdots,\bm{k}_n)=\frac{A_n(M_{\rm h})}{\delta_{\rm c}^n}W(k_1,M_{\rm h})\cdots W(k_n,M_{\rm h})
\nonumber\\
&\quad+\frac{A_{n-1}(M_{\rm h})\sigma^n_M}{\delta_{\rm c}^n}
\frac{d}{d\ln{\sigma_M}}\left[\frac{W(k_1,M_{\rm h})\cdots W(k_n,M_{\rm h})}{\sigma^n_M}\right]\,,
\label{cnl-wind}\\
& {\rm with\;\;\;\;} A_n(M_{\rm h})\equiv\sum^n_{j=0}\frac{n!}{j!}\delta_{\rm c}^j\,
(-\sigma_M)^{-j}f_{\rm MF}^{-1}(\nu)\frac{d^jf_{\rm MF}(\nu)}{d\nu^j}. 
\end{align}
%%%%%%%%%%%%%%%%%%%%%%%%%%%%%%%%%%%%%%%%%%%%%%%%%%%%%%%%%%%%%%%%%%%%%%%%%%%

Here the quantity $\delta_{\rm c}$
is the so-called critical density of the spherical collapse model  whose numerical value is  $\delta_{\rm c} \simeq 1.68$, 
$W(k,M_{\rm h})$ is the top-hat window function over mass scale $R=(3M_{\rm h}/4\pi\rho_{\rm m})^{1/3}$, $M_{\rm h}$ is the halo mass and $\rho_{\rm m}$ is the matter density. 
The quantity $\sigma_{\rm M}$ is the dispersion of smoothed matter density field over mass scale $R$:
%%%%%%%%%%%%%%%%%%%%%%%%%%%%%%%%%%%%%%%%%%%%%%%%%%%%%%%%%%%%%%%%%%%%%%%%%%%
\begin{align}
\sigma_{\rm M}^2=\int\frac{k^2dk}{2\pi^2}W^2(k,M_{\rm h})P_{\rm L}(k).
\end{align}
%%%%%%%%%%%%%%%%%%%%%%%%%%%%%%%%%%%%%%%%%%%%%%%%%%%%%%%%%%%%%%%%%%%%%%%%%%%
\if0
The function $f_{\rm MF}(\nu)$ is defined through the halo mass function $n(M_{\rm h},z)$:
%%%%%%%%%%%%%%%%%%%%%%%%%%%%%%%%%%%%%%%%%%%%%%%%%%%%%%%%%%%%%%%%%%%%%%%%%%%
\begin{align}
\nu f_{\rm MF}(\nu)&\equiv M_{\rm h}^2\frac{n(M_{\rm h},z)}{\bar{\rho}}\frac{d\log{M_{\rm h}}}{d\log{\nu}}\label{MF},
\end{align}
%%%%%%%%%%%%%%%%%%%%%%%%%%%%%%%%%%%%%%%%%%%%%%%%%%%%%%%%%%%%%%%%%%%%%%%%%%%
where $\nu=\delta_{\rm c}/\sigma_M$. 
\fi
 For  $f_{\rm MF}$ which is a function of $\nu \equiv \delta_{\rm c}/\sigma_M$,
throughout the paper, we adopt the Sheth-Tormen fitting formula for the halo mass function $n(M_{\rm h},z)$ \cite{2001MNRAS.323....1S}, which yields 
%%%%%%%%%%%%%%%%%%%%%%%%%%%%%%%%%%%%%%%%%%%%%%%%%%%%%%%%%%%%%%%%%%%%%%%%%%%
\begin{align}
f_{\rm MF} = f_{\rm ST}(\nu)&=A(p)\sqrt{\frac{2}{\pi}}[1+(q\nu^2)^{-p}]\sqrt{q}\nu e^{-q\nu^2/2},
\label{st}
\end{align}
%%%%%%%%%%%%%%%%%%%%%%%%%%%%%%%%%%%%%%%%%%%%%%%%%%%%%%%%%%%%%%%%%%%%%%%%%%%
 where $A(p)$ is expressed in terms of the Gamma function, $\Gamma(x)$ as  $A(p)=[1+\pi^{-1/2}2^{-p}\Gamma(1/2-p)]^{-1}$
with $p=0.3$, $q=0.707$.

For the matter fluctuation (i.e., $X = {\rm m}$), 
we have $c_n^{\rm L} \simeq 0$, which gives
%%%%%%%%%%%%%%%%%%%%%%%%%%%%%%%%%%%%%%%%%%%%%%%%%%%%%%%%%%%%%%%%%%%%%%%%%%%
\begin{align}
\Gamma^{(1)}_{\rm m}(\bm{k})\simeq 1,~
\Gamma^{(2)}_{\rm m}(\bm{k}_1,\bm{k}_2)\simeq F_2(\bm{k}_1,\bm{k}_2). \label{mattergamma}
\end{align}
%%%%%%%%%%%%%%%%%%%%%%%%%%%%%%%%%%%%%%%%%%%%%%%%%%%%%%%%%%%%%%%%%%%%%%%%%%%

%%%%%%%%%%%%%%%%%%%%%%%%%%%%%%%%%%%%%%%%%%%%%%%%%%%%%%%%%%%%%%%%%%%%%%%%%%%
%%%%%%%%%%%%%%%%%%%%%%%%%%%%%%%%%%%%%%%%%%%%%%%%%%%%%%%%%%%%%%%%%%%%%%%%%%%
\subsection{Angular bispectra in imaging surveys}
%%%%%%%%%%%%%%%%%%%%%%%%%%%%%%%%%%%%%%%%%%%%%%%%%%%%%%%%%%%%%%%%%%%%%%%%%%%
%%%%%%%%%%%%%%%%%%%%%%%%%%%%%%%%%%%%%%%%%%%%%%%%%%%%%%%%%%%%%%%%%%%%%%%%%%%

Based on the three-dimensional bispectra given in the above discussion,
we derive the formulas for angular bispectra projected on the celestial sphere, 
which are statistical quantities observed in imaging surveys.
Employing the flat-sky limit, these statistical quantities are defined as
%%%%%%%%%%%%%%%%%%%%%%%%%%%%%%%%%%%%%%%%%%%%%%%%%%%%%%%%%%%%%%%%%%%%%%%%%%%
\begin{align}
&\frac{1}{3}\Bigl[\left\langle \Delta_{\rm a}(\bm{\ell}_1)\Delta_{\rm b}(\bm{\ell}_2)\Delta_{\rm c}(\bm{\ell}_3)\right\rangle+\mbox{2 perms}\,\,(\bm{\ell}_1\leftrightarrow \bm{\ell}_2\leftrightarrow\bm{\ell}_3)\,\Bigr]
\nonumber\\
&\qquad\qquad\quad\equiv (2\pi)^2\delta_{\rm D}^{(2)}(\bm{\ell}_1+\bm{\ell}_2+\bm{\ell}_3)B_{\rm abc}(\ell_1,\ell_2,\ell_3),
\label{bis2}
\end{align}
%%%%%%%%%%%%%%%%%%%%%%%%%%%%%%%%%%%%%%%%%%%%%%%%%%%%%%%%%%%%%%%%%%%%%%%%%%%
 where  $\delta_{\rm D}^{(2)}$ is the two-dimensional Dirac delta function.
 The quantity $\Delta_{\rm a}$ is the two-dimensional density field projected on the celestial sphere, 
and the subscripts $a$, $b$, $c$ imply either a halo/galaxy number-density fluctuation $\Delta_{\rm h}$ or weak-lensing.  
These are related to the three-dimensional density field
through: 
%%%%%%%%%%%%%%%%%%%%%%%%%%%%%%%%%%%%%%%%%%%%%%%%%%%%%%%%%%%%%%%%%%%%%%%%%%%
\begin{align}
\Delta_{\rm h}(\bm{\theta})&=\int_0^\infty dz~ W_{\rm h}(z)\delta^{(3)}_{\rm h}(\chi(z)\bm{\theta},z),\label{delh}\\
\kappa(\bm{\theta})&=\int_0^\infty dz~ W_\kappa(z)\delta^{(3)}_{\rm m}(\chi(z)\bm{\theta},z),\label{delk}
\end{align}
%%%%%%%%%%%%%%%%%%%%%%%%%%%%%%%%%%%%%%%%%%%%%%%%%%%%%%%%%%%%%%%%%%%%%%%%%%%
 where  $W_{\rm a}$  are  the weight functions given by
%%%%%%%%%%%%%%%%%%%%%%%%%%%%%%%%%%%%%%%%%%%%%%%%%%%%%%%%%%%%%%%%%%%%%%%%%%%
\begin{align}
W_{\rm h}(z)&=\frac{n_{\rm h}(z)}{\bar{n}_{\rm h}}  ,
\\
W_\kappa\left(z\right)&=\frac{4\pi G\rho_{\rm m}\left(z\right)}{H\left(z\right) (1+z)^2 \bar{n}_{\rm s}}\int_z^\infty dz'~n_{\rm s}(z')\frac{\left(\chi\left(z'\right)-\chi\left(z\right)\right)\chi\left(z\right)}{\chi\left(z'\right)}.\label{weightkappa}
\end{align}
%%%%%%%%%%%%%%%%%%%%%%%%%%%%%%%%%%%%%%%%%%%%%%%%%%%%%%%%%%%%%%%%%%%%%%%%%%%
Here, we denote
the projected number density of halo by $\bar{n}_{\rm h}$ and its redshift distribution per unit area by $n_{\rm h}(z)$.  These quantities are 
respectively given by
%%%%%%%%%%%%%%%%%%%%%%%%%%%%%%%%%%%%%%%%%%%%%%%%%%%%%%%%%%%%%%%%%%%%%%%%%%%
\begin{align}
\bar{n}_{\rm h}=\int_0^\infty dz~n_{\rm h}(z)=\int_0^\infty dz\,\frac{\chi^2(z)}{H(z)}\int_{M_{\rm min}}^\infty dM_{\rm h}~n(M_{\rm h},z),
\label{nlnl}
\end{align}
%%%%%%%%%%%%%%%%%%%%%%%%%%%%%%%%%%%%%%%%%%%%%%%%%%%%%%%%%%%%%%%%%%%%%%%%%%%
where  $\chi$ is  the comoving radial distance and $n(M_{\rm h},z)$ is the halo mass function.
  $M_{\rm min}$ is the minimum mass of observed halos, where we set $M_{\rm min}$ to $10^{13}~h^{-1}M_{\odot}$ through this paper.
For the redshift distribution of source galaxies for weak-gravitational lensing observations, denoted by $n_{\rm s} (z)$,
we adopt the following functional form (e.g., \cite{2011PhRvD..83l3514N}): 
%%%%%%%%%%%%%%%%%%%%%%%%%%%%%%%%%%%%%%%%%%%%%%%%%%%%%%%%%%%%%%%%%%%%%%%%%%%
\begin{align}
 n_{\rm s}(z)dz=\bar{n}_{\rm s}\frac{3z^2}{2(0.64z_{\rm m})^3}\exp{\left[-\left(\frac{z}{0.64z_{\rm m}}\right)^{3/2}\right]}dz\, .\label{keiken}
\end{align}
%%%%%%%%%%%%%%%%%%%%%%%%%%%%%%%%%%%%%%%%%%%%%%%%%%%%%%%%%%%%%%%%%%%%%%%%%%%
Employing the Limber approximation\footnote{ It has been known that the Limber approximation becomes invalid at the large-angular scales.
However, 
as shown in Ref. \cite{2008PhRvD..78l3506L},
in the case of a wide observed redshift range, 
the Limber approximation can be applied even at large-angular scales.
In our analysis, we mainly investigate the cases with single-redshift bin, 
and hence our expression based on the Limber approximation should not be invalid.}
 \cite{1954ApJ...119..655L} which is valid in the flat-sky limit, we finally obtain
%%%%%%%%%%%%%%%%%%%%%%%%%%%%%%%%%%%%%%%%%%%%%%%%%%%%%%%%%%%%%%%%%%%%%%%%%%%
\begin{align}
&B_{\rm abc}(\bm{\ell}_1,\bm{\ell}_2,\bm{\ell}_3)=\int dz\frac{H^2(z)}{\chi^4\left(z\right)} W_{\rm a}\left(z\right)W_{\rm b}\left(z\right)W_{\rm c}\left(z\right) 
\nonumber\\
&\qquad\qquad\qquad\qquad\times
\mathcal{B}_{\rm XYZ}\left(\frac{\bm{\ell}_1}{\chi\left(z\right)},\frac{\bm{\ell}_2}{\chi\left(z\right)},\frac{\bm{\ell}_3}{\chi\left(z\right)};\,z\right).\label{blimber}
\end{align}
%%%%%%%%%%%%%%%%%%%%%%%%%%%%%%%%%%%%%%%%%%%%%%%%%%%%%%%%%%%%%%%%%%%%%%%%%%%

%%%%%%%%%%%%%%%%%%%%%%%%%%%%%%%%%%%%%%%%%%%%%%%%%%%%%%%%%%%%%%%%%%%%%%%%%%%
%%%%%%%%%%%%%%%%%%%%%%%%%%%%%%%%%%%%%%%%%%%%%%%%%%%%%%%%%%%%%%%%%%%%%%%%%%%
\section{Forecast constraints on equilateral-type primordial non-Gaussianity\label{sec:fifi}}
%%%%%%%%%%%%%%%%%%%%%%%%%%%%%%%%%%%%%%%%%%%%%%%%%%%%%%%%%%%%%%%%%%%%%%%%%%%
%%%%%%%%%%%%%%%%%%%%%%%%%%%%%%%%%%%%%%%%%%%%%%%%%%%%%%%%%%%%%%%%%%%%%%%%%%%

In this section, 
based on the Fisher matrix formalism, 
let us quantitatively estimate the impact of the bispectra and weak-gravitational lensing effect
on the observational constraints on the equilateral-type primordial non-Gaussianity. 
Here, as representative future/ongoing imaging surveys, we shall consider three representative surveys: the Subaru Hyper Suprime-Cam (HSC)~\cite{HSCrev}, the Dark Energy Survey (DES)~\cite{2005astro.ph.10346T}, and the Large Synoptic Survey Telescope (LSST)~\cite{2009arXiv0912.0201L}. 
Imaging surveys are characterized by the survey area $f_{\rm sky}\equiv\Omega_{\rm s}/4\pi$, the mean source redshift $z_{\rm m}$, and the mean number density of source galaxies per unit area $\bar{n}_{\rm s}$. 
 We take the values of these parameters for the representative surveys 
as $(f_{\rm sky},~z_{\rm m},~\bar{n}_{\rm s} [{\rm arcmin}^{-2}]) = (0.0375~(1500\,{\rm deg}^2),~1.0,~35)$ for HSC~\cite{HSCrev}, $(0.125~(5000\,{\rm deg}^2),~0.5,~12)$ for DES~\cite{2005astro.ph.10346T}, and $(0.5~(20000\,{\rm deg}^2),~1.5,~100)$ for LSST~\cite{2009arXiv0912.0201L}.

%%%%%%%%%%%%%%%%%%%%%%%%%%%%%%%%%%%%%%%%%%%%%%%%%%%%%%%%%%%%%%%%%%%%%%%%%%%%%%%%%%%%%
\subsection{Fisher matrix \label{sec:fisher}}
%%%%%%%%%%%%%%%%%%%%%%%%%%%%%%%%%%%%%%%%%%%%%%%%%%%%%%%%%%%%%%%%%%%%%%%%%%%%%%%%%%%%%

Following Ref.~\cite{Hashimoto:2015tnv},
the Fisher matrix for the parameters $\bm{p}$ which characterize the theoretical expression of the angular bispectra $\bm{B}_{i}$ are defined by
%%%%%%%%%%%%%%%%%%%%%%%%%%%%%%%%%%%%%%%%%%%%%%%%%%%%%%%%%%%%%%%%%%%%%%%%%%%
\begin{widetext}
\begin{align}
&F_{\alpha\beta}=\left.\sum^{\ell_{\rm max}}_{\ell_i=\ell_{\rm min}}\frac{\partial\bm{B}_{i}(\bm{p})}{\partial p_\alpha}(\bm{{\rm Cov}}^B)_{ij}^{-1}\frac{\partial\bm{B}_{j}(\bm{p})}{\partial p_\beta}\right|_{\bm{p}=\bm{p}_0},\label{eq:Fisher_B}\\
&\bm{B}_i=\begin{pmatrix}
(B_{\rm hhh})_i \\
(B_{\rm hh\kappa})_i  \\
(B_{\rm h\kappa\kappa})_i  \end{pmatrix},
\qquad
\bm{{\rm Cov}}_{ij}^B=\begin{pmatrix}
{\rm Cov}[(B_{\rm hhh})_i,(B_{\rm hhh})_j]& {\rm Cov}[(B_{\rm hhh})_i,(B_{\rm hh\kappa})_j]& {\rm Cov}[(B_{\rm hhh})_i,(B_{\rm h\kappa\kappa})_j]\\
{\rm Cov}[(B_{\rm hh\kappa})_i,(B_{\rm hhh})_j]& {\rm Cov}[(B_{\rm hh\kappa})_i,(B_{\rm hh\kappa})_j]& {\rm Cov}[(B_{\rm hh\kappa})_i,(B_{\rm h\kappa\kappa})_j]\\
{\rm Cov}[(B_{\rm h\kappa\kappa})_i,(B_{\rm hhh})_j]& {\rm Cov}[(B_{\rm h\kappa\kappa})_i,(B_{\rm hh\kappa})_j]& {\rm Cov}[(B_{\rm h\kappa\kappa})_i,(B_{\rm h\kappa\kappa})_j] \end{pmatrix}\label{covB2}.
\end{align}
\end{widetext}
%%%%%%%%%%%%%%%%%%%%%%%%%%%%%%%%%%%%%%%%%%%%%%%%%%%%%%%%%%%%%%%%%%%%%%%%%%%
Here, $\bm{p}_0$ is a set of fiducial cosmological parameters, subscripts $i$ and $j$ run over all possible triangle configurations
 whose side lengths are within the range $[\ell_{\rm min},\,\ell_{\rm max}]$. 
In our analysis, we set the minimum multipole to $\ell_{\rm min}=\ell_{\rm f}=\sqrt{\pi / f_{\rm sky}}$,
and the maximum multipole $\ell_{\rm max}$ is set to $150$.
Later, we will also discuss the $\ell_{\rm max}$ dependence of the $1\sigma$ errors 
on the non-linearity parameters. 
In the above expression, $\bm{{\rm Cov}}_{ij}^B$ is the angular bispectra covariance matrix and $[{\rm Cov}^B]^{-1}_{ij}$
is its inverse. 
Then, 
assuming the Gaussian covariances\footnote{ 
In our analysis, we set the maximum multipole $\ell_{\rm max}$ to be $150$,
and it was shown that for $\ell\lesssim 200$ the assumption of Gaussian covariance matrices is expected to be valid \cite{2013MNRAS.429..344K}.
For $\ell \gtrsim 200$,
 the non-linear evolution of the matter density field does not become negligible, and we need to take into account 
 the gravity-induced non-Gaussian contribution.}
, which is based on the fact that large primordial non-Gaussianity is not allowed by the current observations,
the covariance matrix of the angular bispectra for a given set of multipole bins $(\ell_i,\,\ell_j,\,\cdots)$ 
can be given by \cite{2013MNRAS.429..344K}
%%%%%%%%%%%%%%%%%%%%%%%%%%%%%%%%%%%%%%%%%%%%%%%%%%%%%%%%%%%%%%%%%%%%%%%%%%%
\begin{align}
&{\rm Cov}[B_{\rm abc}(\ell_i,\ell_j,\ell_k),B_{\rm a'b'c'}(\ell_l,\ell_m,\ell_n)]
=\frac{1}{9}\frac{\Omega_{\rm s}}{N_{\rm trip}(\ell_i,\ell_j,\ell_k)}
\nonumber\\
&\times\Biggl[\Bigl\{\left(C_{\rm aa'}(\ell_i)+N_{\rm aa'}\right)\left(C_{\rm bb'}(\ell_j)+N_{\rm bb'}\right)\left(C_{\rm cc'}(\ell_k)+N_{\rm cc'}\right)
\nonumber\\
&\quad\times\left(\delta^{\rm K}_{\bm{\ell}_i+\bm{\ell}_l}\delta^{\rm K}_{\bm{\ell}_j+\bm{\ell}_m}\delta^{\rm K}_{\bm{\ell}_k+\bm{\ell}_n}+\mbox{5 perms}\,\,(\ell_l\leftrightarrow \ell_m\leftrightarrow \ell_n)\right)
\nonumber\\
&\quad+\mbox{2 perms}\,\,({\rm a'\leftrightarrow b'\leftrightarrow c'})\Bigr\}
+\mbox{2 perms}\,\,({\rm a\leftrightarrow b\leftrightarrow c})\Biggr],
\label{covB}
\end{align}
%%%%%%%%%%%%%%%%%%%%%%%%%%%%%%%%%%%%%%%%%%%%%%%%%%%%%%%%%%%%%%%%%%%%%%%%%%%
with $\delta^{\rm K}_{\bm{\ell}_i+\bm{\ell}_j}$ being the Kronecker delta.
Here, 
$N_{\rm ab}$ is the shot-noise contribution, given by $N_{\rm ab} = 1/\bar{n}_{\rm h}$ (${\rm ab} = {\rm hh}$), $\sigma_\gamma / \bar{n}_s$ (${\rm ab} = \kappa\kappa$)
and $0$ (otherwise).
$C_{\rm ab}$ in the above expression is the angular power spectra of halo/galaxy clustering and weak-gravitational lensing.
Note that, here, we calculated the angular power spectra
by using the Limber approximation, as is the case in Eq. (\ref{blimber}).
The quantity $\sigma_\gamma$ represents the dispersion of the intrinsic shape noise and we adopt $\sigma_\gamma=0.3$ \cite{2013PhR...530...87W}. 
$N_{\rm trip}(\ell_i,\ell_j,\ell_k)$ is the number of independent triplets that form a  triangular configuration respectively within the $i$-, $j$-, and $k$-th bins. 
In the limit $\ell_i\gg\ell_{\rm f} = \ell_{\rm min}$, 
we obtain a simple analytical expressions for $N_{\rm trip}$ \cite{2013MNRAS.429..344K}: 
%%%%%%%%%%%%%%%%%%%%%%%%%%%%%%%%%%%%%%%%%%%%%%%%%%%%%%%%%%%%%%%%%%%%%%%%%%%
\begin{align}
&N_{\rm trip}(\ell_i,\ell_j,\ell_k)\simeq 2\frac{(2\pi \ell_i\Delta \ell_i)(\ell_j\Delta\varphi_{12}\Delta \ell_j)}{\ell_{\rm f}^4}
\end{align}
%%%%%%%%%%%%%%%%%%%%%%%%%%%%%%%%%%%%%%%%%%%%%%%%%%%%%%%%%%%%%%%%%%%%%%%%%%%
 where the angle $\Delta\varphi_{12}$ is given by
%%%%%%%%%%%%%%%%%%%%%%%%%%%%%%%%%%%%%%%%%%%%%%%%%%%%%%%%%%%%%%%%%%%%%%%%%%%
\begin{align}
\Delta\varphi_{12}(\ell_i,\ell_j,\ell_k)&\simeq (\sin{\varphi_{12}})^{-1}\frac{\ell_k\Delta \ell_k}{\ell_i\ell_j}
\nonumber
\\
&=\frac{2\ell_k\Delta \ell_k}{\sqrt{2\ell_i^2\ell_j^2+2\ell_i^2\ell_k^2+2\ell_j^2\ell_k^2-\ell_i^4-\ell_j^4-\ell_k^4}}. 
\label{ntrip}
\end{align}
%%%%%%%%%%%%%%%%%%%%%%%%%%%%%%%%%%%%%%%%%%%%%%%%%%%%%%%%%%%%%%%%%%%%%%%%%%%
 Note that the width of the $i$-th multipole bin should be larger than the minimum multipole, i.e., $\Delta \ell_i> \ell_{\rm f}$.

\subsection{$1\sigma$ errors on the equilateral-type non-Gaussianities}

Here, for free parameters $\bm{p}$, we consider the two equilateral-type non-Gaussian parameters, i.e., $\bm{p}=(f_{\rm NL}^{\rm equil}, \,g_{\rm NL}^{(\partial\sigma)^4})$, with the fiducial values of $\bm{p}_0=(0,0)$. Note that we do not marginalize the uncertainty in the halo bias properties,
since in iPT we can completely specify the halo bias by fixing the mass of observed halos. 

%%%%%%%%%%%%%%%%%%%%%%%%%%%%%%%%%%%%%%%%%%%%%%%%%%%%%%%%%%%%%%%%%%%%%%%%%%%
\begin{figure*}[htbp]
  \begin{center}
    \begin{tabular}{c}

      % 1
      \begin{minipage}{0.33\hsize}
        \begin{center}
          \includegraphics[width=55mm,angle=270]{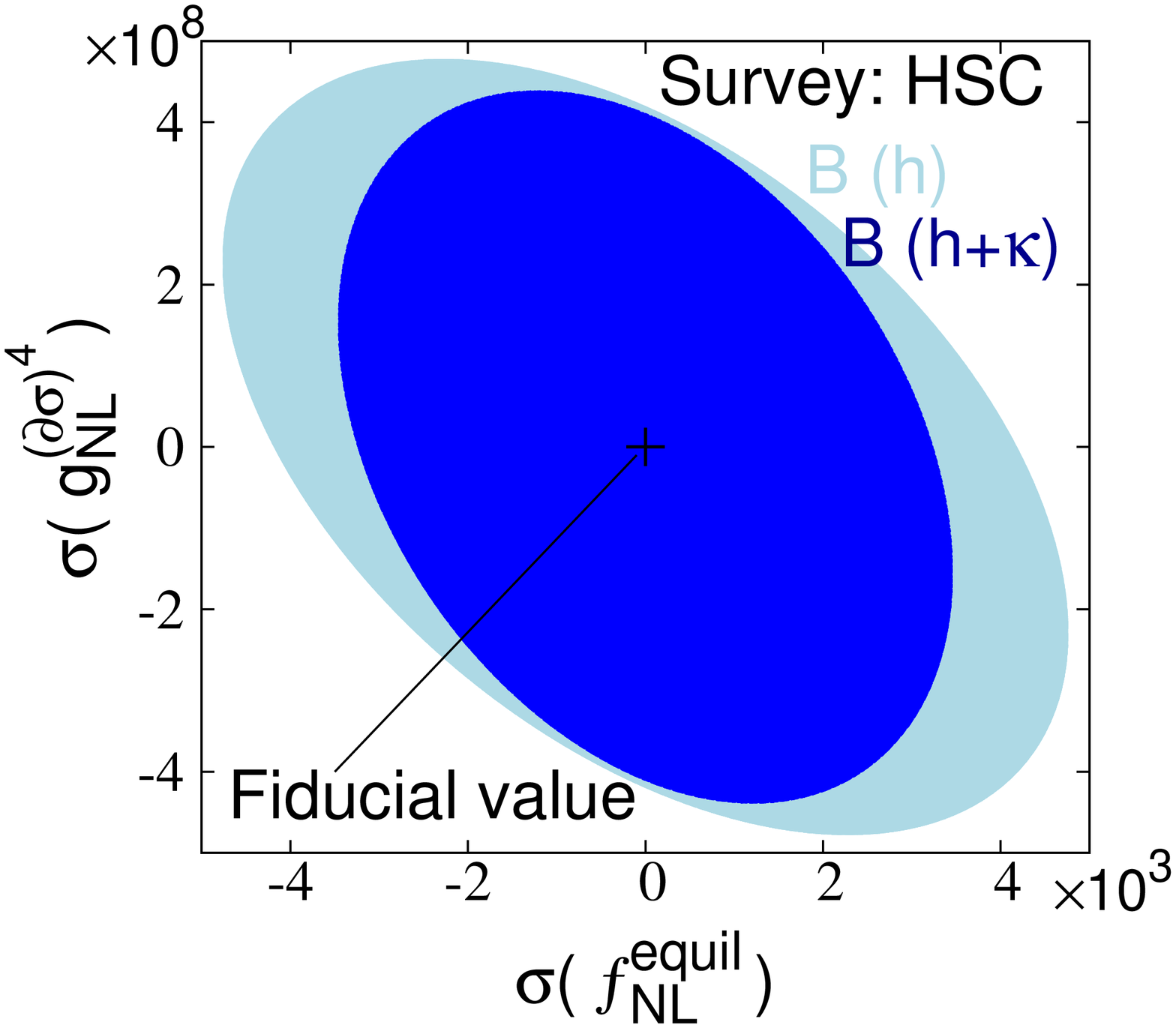}
        \end{center}
       \end{minipage}

      % 2
      \begin{minipage}{0.33\hsize}
        \begin{center}
          \includegraphics[width=55mm,angle=270]{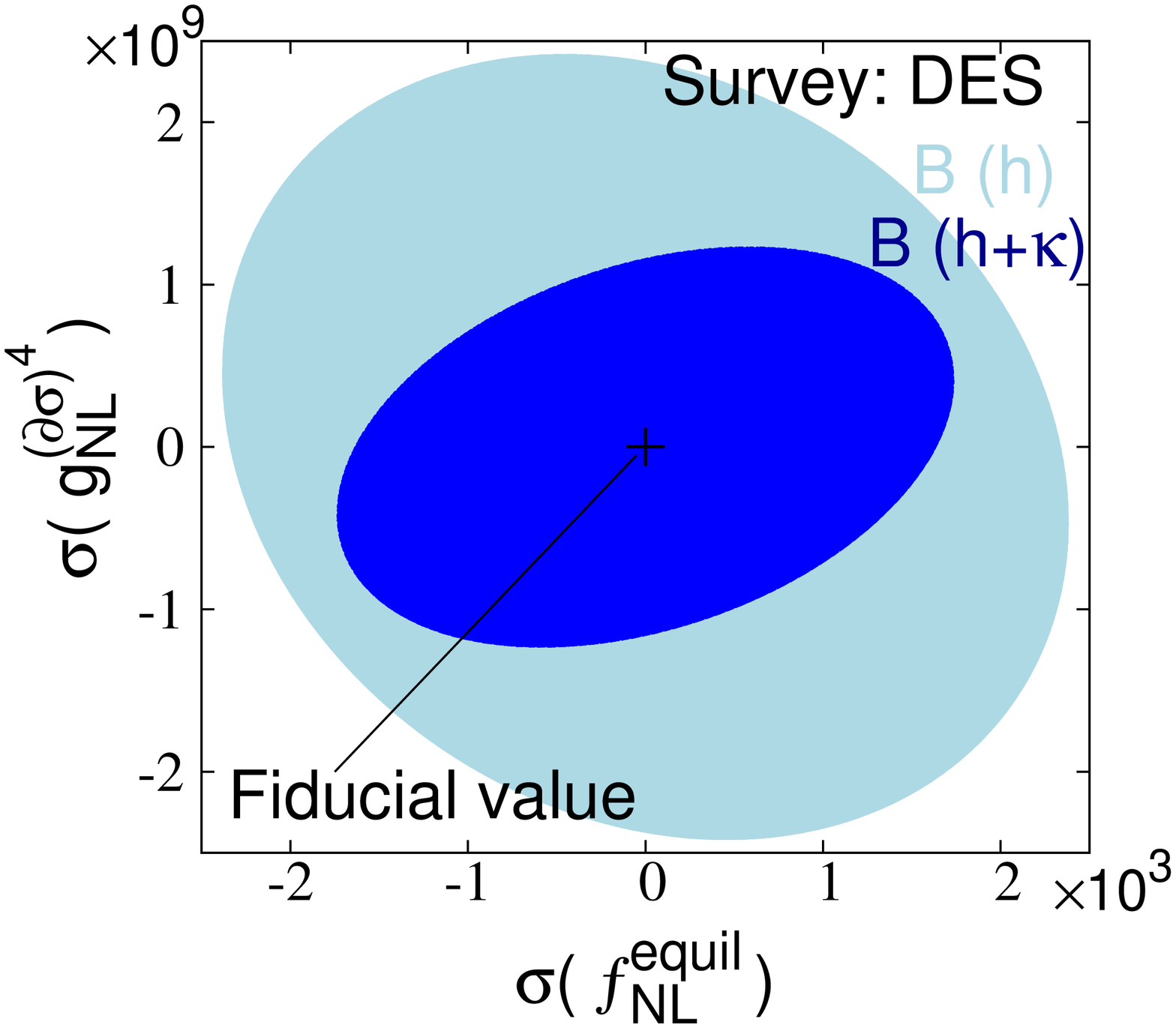}
        \end{center}
      \end{minipage}

      % 3
      \begin{minipage}{0.33\hsize}
        \begin{center}
          \includegraphics[width=55mm,angle=270]{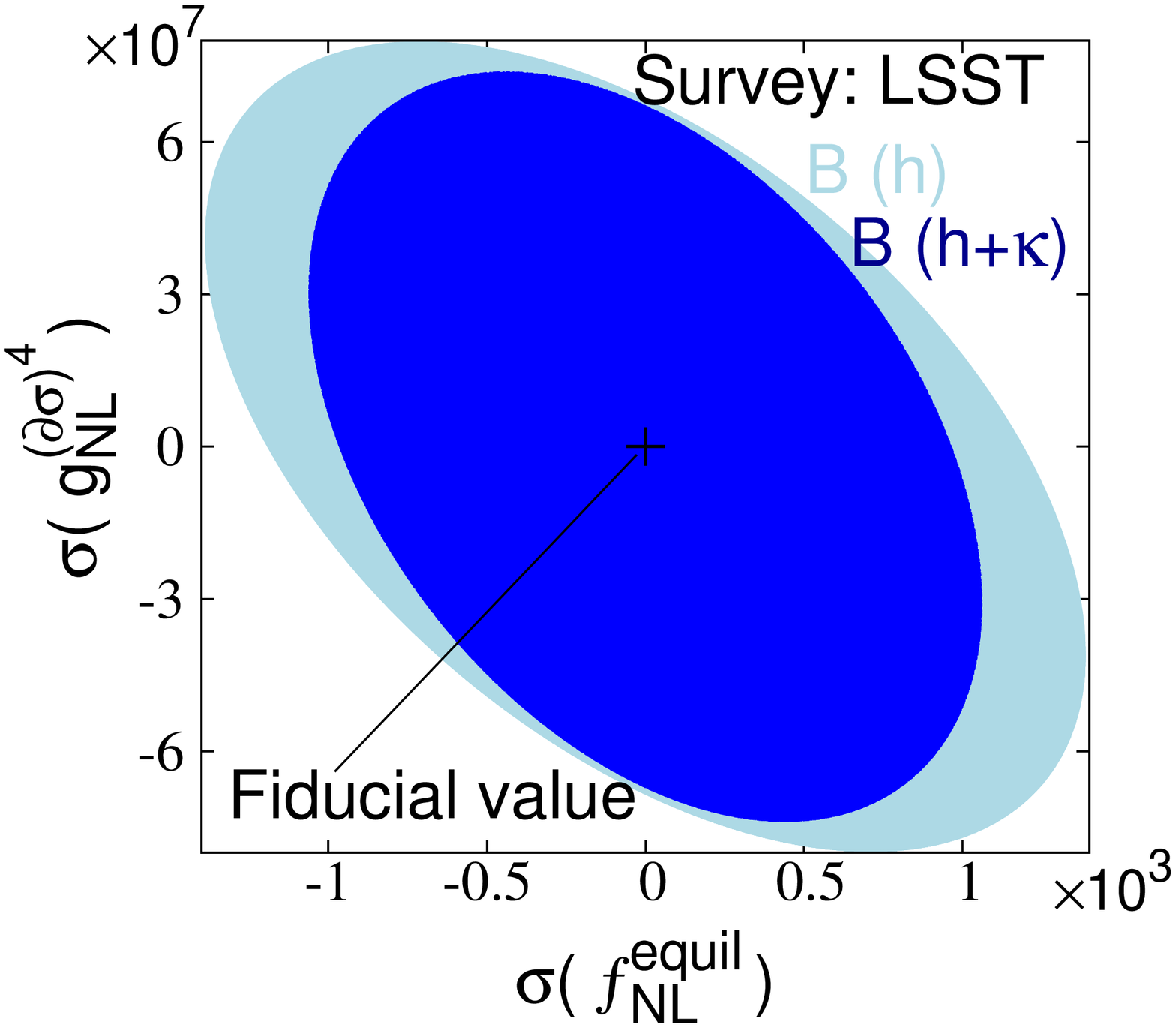}
        \end{center}
      \end{minipage}

    \end{tabular}
      \end{center}
      
        \caption{Forecast results of equilateral-type primordial non-Gaussian parameters by HSC (left), DES (middle), and LSST (right). In each panel, marginalized $1\sigma$ error contours on $g_{\rm NL}^{(\partial\sigma)^4}-f_{\rm NL}^{\rm equil}$ planes are shown. Light blue contours represent the constraints from the auto-angular bispectrum of halo/galaxy clustering, while the blue contours indicate the constraints when we add cross-angular bispectra between halos and weak lensing. 
Notice that the plotted range of the error contours is changed in each panel, for clarity. 
\label{const_LSSTDESHSC} }
\end{figure*}
%%%%%%%%%%%%%%%%%%%%%%%%%%%%%%%%%%%%%%%%%%%%%%%%%%%%%%%%%%%%%%%%%%%%%%%%%%%

%%%%%%%%%%%%%%%%%%%%%%%%%%%%%%%%%%%%%%%%%%%%%%%%%%%%%%%%%%%%%%%%%%%%%%%%%%%   
  %%%%%%%%%%%%%%%%%%%%%%%%%%%%%%%%%%%%%%%%%%%%%%%%%%%%%%%%%%%%%%%%%%%%%%%%%%%
\begin{table*}[htbp]
\begin{ruledtabular}
  \begin{tabular}{c|ll|lll}
     Survey&&&$B_{\rm hhh}$&$B_{\rm hhh}+B_{\rm hh\kappa}+B_{\rm h\kappa\kappa}$&\\ \hline
    HSC&$\sigma (f_{\rm NL}^{\rm equil})$&&$3.2\times 10^3\ (2.9\times 10^3)$&$2.3\times 10^3\ (2.1\times 10^3)$&\\ 
    &$\sigma (g_{\rm NL}^{(\partial\sigma)^4})$&&$3.2\times 10^8\ (2.9\times 10^8)$&$2.9\times 10^8\ (2.7\times 10^8)$&\\ 
 \hline   
  DES&$\sigma (f_{\rm NL}^{\rm equil})$&&$1.6\times 10^3\ (1.6\times 10^3)$&$1.1\times 10^3\ (1.1\times 10^3)$&\\ 
    &$\sigma (g_{\rm NL}^{(\partial\sigma)^4})$&&$1.6\times 10^{9}\ (1.7\times 10^9)$&$8.2\times 10^{8}\ (7.7\times 10^8)$&\\     
     \hline
     LSST&$\sigma (f_{\rm NL}^{\rm equil})$&&$9.2\times 10^2\ (8.0\times 10^2)$&$7.0\times 10^2\ (6.4\times 10^2)$&\\ 
    &$\sigma (g_{\rm NL}^{(\partial\sigma)^4})$&&$5.3\times 10^7\ (4.6\times 10^7)$&$4.9\times 10^{7}\ (4.4\times 10^{7})$&\\   
    \end{tabular}
  \caption{Forecast results of marginalized (un-marginalized) $1\sigma$ errors on equilateral type primordial non-Gaussian parameters
, $\fnleq$ and $\gnl^{(\partial \sigma)^4}$ for HSC, DES, and LSST.}
  \label{hhhdddlll}
  \end{ruledtabular}
  \end{table*}
%%%%%%%%%%%%%%%%%%%%%%%%%%%%%%%%%%%%%%%%%%%%%%%%%%%%%%%%%%%%%%%%%%%%%%%%%%%

%%%%%%%%%%%%%%
The results of Fisher matrix analysis are summarized in Fig.~\ref{const_LSSTDESHSC}, and Tab. \ref{hhhdddlll}. The elliptic contours in Fig.~\ref{const_LSSTDESHSC} show the marginalized $1\sigma$ error constraints on $\fnleq$ and $\gnl^{(\partial \sigma)^4}$. 
Light blue contours represent the constraints from the auto-angular bispectrum of halo/galaxy clustering, while the blue contours indicate the constraints when we add cross-angular bispectra between halos and weak lensing. 
The degeneracy of these two parameters is not very strong because the contributions from these parameters have different shape dependences \cite{2015PhRvD..91l3521M}. Moreover, including the cross-angular  bispectra partly breaks the parameters degeneracy, which makes the constraints tighter in each survey. In particular, the constraints by DES turn out to be most improved by taking into account the cross-angular  bispectra.
As we see in Tab. \ref{hhhdddlll}, the tightest constraints in the three surveys come from LSST on both $f_{\rm NL}^{\rm equil}$ and $g_{\rm NL}^{(\partial\sigma)^4}$. Interestingly, the constraint on $f_{\rm NL}^{\rm equil}$ by DES is tighter than that by HSC, on the other hand, the constraint on $g_{\rm NL}^{(\partial\sigma)^4}$ is opposite. This result implies that deep imaging surveys are advantageous to give tighter constraints on the equilateral-type primordial trispectrum, denoted by  $g_{\rm NL}^{(\partial\sigma)^4}$,
while wide imaging surveys are advantageous for the equilateral-type primordial bispectrum, denoted by $f_{\rm NL}^{\rm equil}$.

%%%%%%%%%%%%%%%%%%%%%%%%%%%%%%%%%%%%%%%%%%%%%%%%%%%%%%%%%%%%%%%%%%%%%%%%%%%
\begin{figure*}[htbp]
  \begin{center}
    \begin{tabular}{c}

      % 1
      \begin{minipage}{0.5\hsize}
        \begin{center}
          \includegraphics[width=55mm,angle=270]{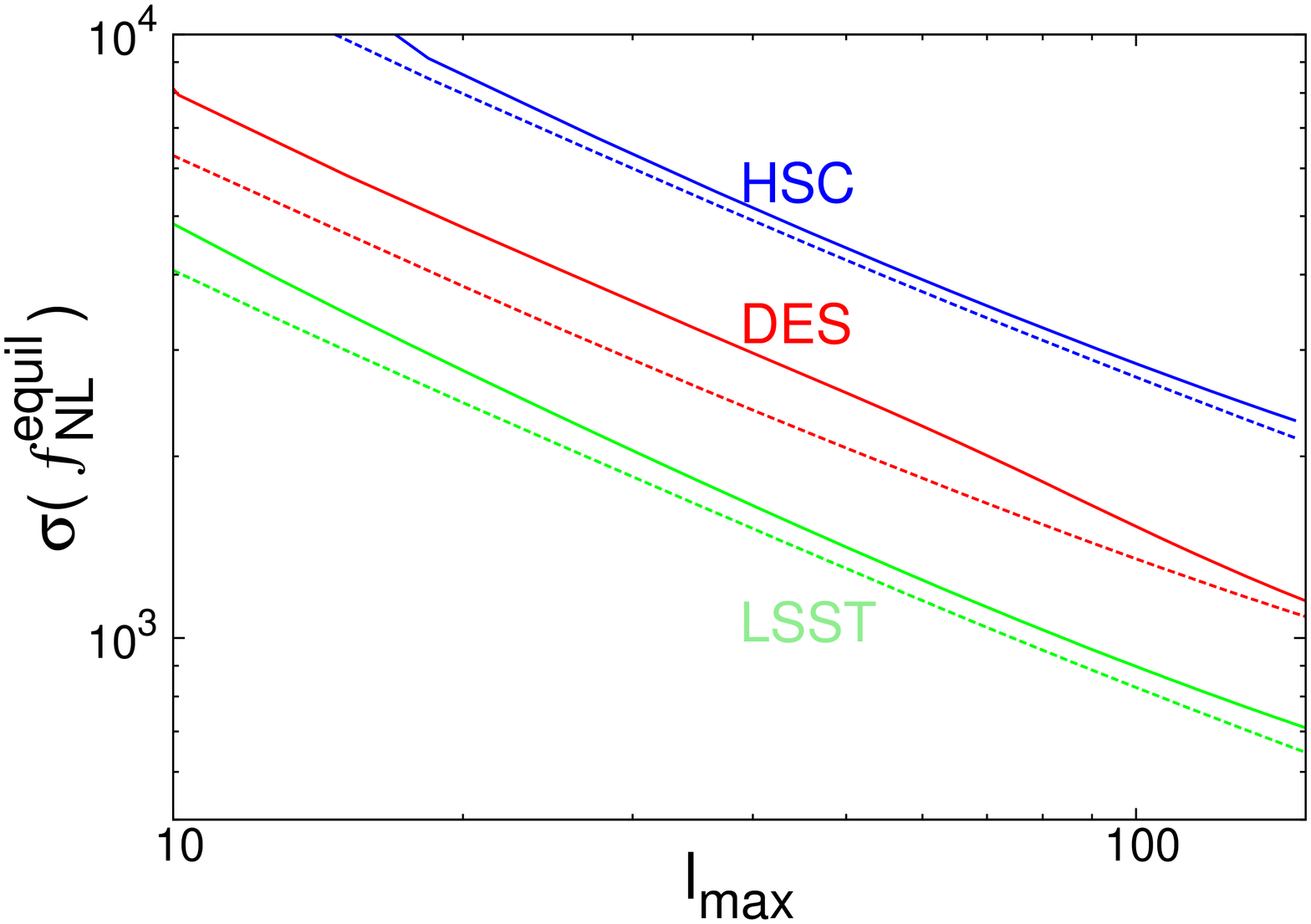}
        \end{center}
      \end{minipage}

      % 2
      \begin{minipage}{0.5\hsize}
        \begin{center}
          \includegraphics[width=55mm,angle=270]{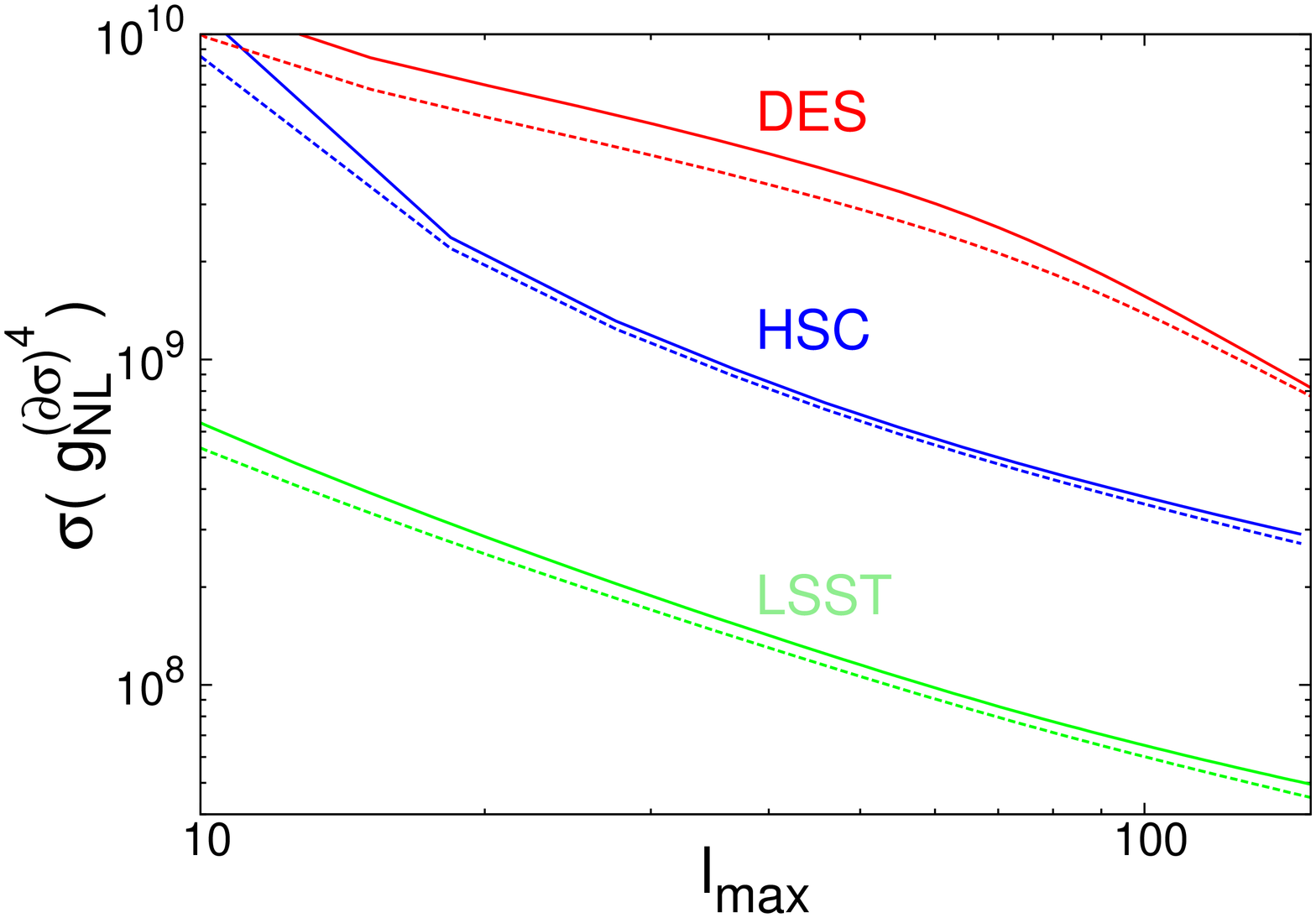}
        \end{center}
        \end{minipage}

    \end{tabular}
     \end{center}
       \caption{Sensitivity of the marginalized $1\sigma$ errors on $f_{\rm NL}^{\rm equil}$ (left), $g_{\rm NL}^{(\partial\sigma)^4}$ (right) to the parameters $\ell_{\rm max}$. Solid (dashed) lines represent marginalized (un-marginalized) $1\sigma$ errors on equilateral type primordial non-Gaussian parameters for HSC (blue), DES (red), and LSST (green).
 \label{lmaxdep_const}}
\end{figure*}
%%%%%%%%%%%%%%%%%%%%%%%%%%%%%%%%%%%%%%%%%%%%%%%%%%%%%%%%%%%%%%%%%%%%%%%%%%%

Before summarizing,
we would like to mention the dependence of the resultant constraints on 
the maximum multipoles, $\ell_{\rm max}$.
While in the above result we have fixed $\ell_{\rm max}$ to be $150$,
in Fig.~\ref{lmaxdep_const} we plot the constraints on $f_{\rm NL}^{\rm equil}$ and $g_{\rm NL}^{(\partial\sigma)^4}$ by HSC, DES and LSST as functions of $\ell_{\rm max}$. 
As can be seen in this figure,
the constraints coming from the bispectra continuously become improved as we increase $\ell_{\rm max}$, because of the increased number of usable triangles. 
However, for a larger value of $\ell_{\rm max}$, 
we need to take into account the late time non-linear evolution more seriously.
We leave more precise forecasts including such a non-linear effect to future work.

%%%%%%%%%%%%%%%%%%%%%%%%%%%%%%%%%%%%%%%%%%%%%%%%%%%%%%%%%%%%%%%%%%%%%%%%%%%%%%%%%%%%%
%%%%%%%%%%%%%%%%%%%%%%%%%%%%%%%%%%%%%%%%%%%%%%%%%%%%%%%%%%%%%%%%%%%%%%%%%%%%%%%%%%%%%  
\section{Summary and Discussion\label{sec:sum}}
%%%%%%%%%%%%%%%%%%%%%%%%%%%%%%%%%%%%%%%%%%%%%%%%%%%%%%%%%%%%%%%%%%%%%%%%%%%%%%%%%%%%%
%%%%%%%%%%%%%%%%%%%%%%%%%%%%%%%%%%%%%%%%%%%%%%%%%%%%%%%%%%%%%%%%%%%%%%%%%%%%%%%%%%%%%

In this paper, we have investigated the impact of angular bispectra from future imaging surveys to obtain constraints on the equilateral-type primordial non-Gaussianities. 
As non-linearity parameters characterizing such non-Gaussianities,
we focus on $\fnleq$ and $\gnl^{(\partial \sigma)^4}$ and obtain simultaneous constraints on these two parameters.
The parameter $\gnl^{(\partial \sigma)^4}$ is related to one of the three equilateral-type trispectra, where the constraints were obtained by the CMB observations  \cite{Smith:2015uia}.
We have neglected the other two $\gnleq$ parameters because it had been shown that one of them, $ g_{\rm NL} ^{\dot{\sigma}^4}$, could not be constrained from the bispectrum and 
the other, $g_{\rm NL} ^{\dot{\sigma}^2 (\partial \sigma)^2}$, has similar shape dependence for $\gnl^{(\partial \sigma)^4}$ in \cite{2015PhRvD..91l3521M}.  

By using the integrated perturbation theory (iPT), we can systematically incorporate both the non-Gaussian mode-coupling from primordial polyspectra and non-linear halo biasing into theoretical template of bispectra. Therefore, we have employed this method to estimate the constraints on $\fnleq$ and $\gnl^{(\partial \sigma)^4}$ by the Fisher matrix analysis. As a result, we have shown that bispectra can give the constraints on $\fnleq$ and $\gnl^{(\partial \sigma)^4}$ for the three representative surveys (HSC, DES and LSST), even though power spectra can not give the constraints on the equilateral-type primordial non-Gaussianity. We have also shown that by combining weak lensing data, the constraints on $\fnleq$ and $\gnl^{(\partial \sigma)^4}$ are improved. In particular, in the case of DES, the constraint on $\gnl^{(\partial \sigma)^4}$ becomes almost twice as tight by combining weak lensing data. The tightest constraints come from LSST, and its expected $1\sigma$ errors on $\fnleq$ and $\gnl^{(\partial \sigma)^4}$ are respectively given by $7.0\times 10^2$ and $4.9\times 10^7$. 
The resultant constraints for all three surveys are somewhat looser than the ones from the current CMB observations \cite{Smith:2015uia, 2015arXiv150201592P}. 
Regardless of this, they could be used for consistency check of CMB observations. In addition, the results forecast in this paper may be a bit conservative because we have considered only the large-angular scale of $\ell\le \ell_{\rm max}=150$. As we have shown in Fig.~\ref{lmaxdep_const}, increasing maximum multipoles,  $\ell_{\rm max}$,
makes the constraints tighter, due to the increased number of triangles which we can use. To roughly estimate the impact of using higher $\ell_{\rm max}$, we calculated the constraints by using $\ell_{\rm max}=400$, and we obtained $\sigma (f_{\rm NL}^{\rm equil})=4.6\times 10^2(3.7\times 10^2)$ and $\sigma (g_{\rm NL}^{(\partial\sigma)^4})=2.9\times10^7(2.3\times10^7)$, as the results of $1\sigma$ marginalized (un-marginalized) errors from LSST. The constraints monotonically decrease for larger $\ell_{\rm max}$ at least $\ell_{\rm max}<400$ in our analysis.
 However, on small-angular scales, the higher-order contribution from gravitational evolution and non-Gaussian feature of error covariance which we have neglected in this paper become important, and a more careful study is necessary. Also, tomographic techniques gives us another chance to improve our constraints on the equilateral-type primordial non-Gaussianity. 
In a previous paper \cite{Hashimoto:2015tnv}, two of us simply estimated the impact of tomographic techniques in the case of local-type primordial non-Gaussianity, and the constraints were improved by a factor of $1.4$ to $3$. We expect similar amount of improvement in the case of equilateral-type primordial non-Gaussianity.

Our analysis in this paper has been based on predictions with iPT, assuming a prior knowledge of halo bias properties. This treatment is consistent with previously known analytic treatment, and thus our results are qualitatively correct. However, for a practical application, further quantitative study is necessary, because the prediction of bispectrum based on iPT with non-Gaussian initial conditions has not been tested against the halo clustering in $N$-body simulations. In addition, for a proper comparison with observations, we need to incorporate nuisance parameters into the characterization of halo bias to reduce the impact of unknown systematics.

At the level of real observation, several sources of systematics (i.e., survey geometry and mask) can arise. It is known that such kind of systematics can mimic primordial non-Gaussianity in the case of power spectrum analysis \cite{2013MNRAS.432.2945H}. Therefore, investigating the effect of such contaminations on bispectra and improving calibration schemes can be important future work for application of bispectrum analysis for photometric surveys. 

\begin{figure*}[htbp]
  \begin{center}
    \begin{tabular}{c}
          \includegraphics[width=70mm,angle=270]{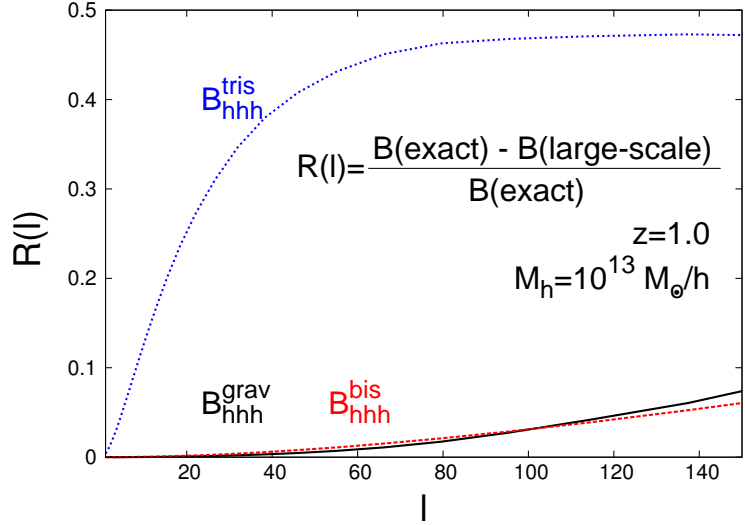}
    \end{tabular}
     \end{center}
       \caption{Relative errors between exact and large-scale limit formulas of the auto-angular bispectrum of halo/galaxy clustering. Each line shows the relative errors of individual contributions of bispectrum from non-linear gravitational evolution (black-solid), primordial bispectrum (red-dashed) and primordial trispectrum (blue-dotted). For simplicity, redshift $(z=1)$ and halo mass $(M_{\rm h}=10^{13}M_\odot/h)$ are fixed.  \label{relative}}
\end{figure*}

Finally, we need to comment about the validity of the large-scale limit approximation used to calculate Eqs.~(\ref{treegrav}), (\ref{Btreebis-L}) and (\ref{1-loopgnlB}). Fig. \ref{relative} shows the relative errors between the exact and large-scale limit formulas of the auto-angular bispectrum of halo/galaxy clustering. Here, the redshift and halo mass are fixed at $z=1$ and $M_{\rm h}=10^{13}~M_{\odot}/h$, and the integration of the redshift and halo mass in Eq.~(\ref{blimber}) are removed for simplicity. The relative errors of the contributions from non-linear gravitational evolution and $\fnleq$ are smaller than $10\%$ even at $\ell=150$. However, even in low $\ell$, the relative error of the contribution from $\gnl^{(\partial \sigma)^4}$ becomes large, and the exact formula is almost twice larger than the large-scale limit formula. Therefore our resultant constraints on $\gnl^{(\partial \sigma)^4}$ could be modified by a factor. Since
our analysis was ultimately not very strict as it contained many approximations (e.g. the Limber approximation and Gaussian covariance matrix), we think that
the errors caused by the large-scale limit are comparable with the ones caused by using other approximations.
Once the importance of tighter constraints on the equilateral-type primordial trispectrum through the halo/galaxy bispectrum is recognized in the future,
we will come back to this topic again and explore it without relying the approximations mentioned above.

%%%%%%%%%%%%%%%%%%%%%%%%%%%%%%%%%%%%%%%%%%%%%%%%%%%%%%%%%%%%%%%%%%%%%%%%%%%
%%%%%%%%%%%%%%%%%%%%%%%%%%%%%%%%%%%%%%%%%%%%%%%%%%%%%%%%%%%%%%%%%%%%%%%%%%%

\acknowledgments

This work was supported in part by JSPS Grants-in-Aid for Scientific
Research, Grants No. 15K17659(S.Y.), No. 15H05888 (S.Y.), No. 16H01103 (S.Y.), and No. 16K17709 (S. M.). We would like to thank Atsushi Taruya, Takahiko Matsubara, Toshiya Namikawa, and Daisuke Yamauchi for useful discussions.

\newpage

\bibliographystyle{JHEP}
\bibliography{bib2}

\providecommand{\href}[2]{#2}\begingroup\raggedright\begin{thebibliography}{10}

\bibitem{Bartolo:2004if}
N.~Bartolo, E.~Komatsu, S.~Matarrese, and A.~Riotto, {\it {Non-Gaussianity from
  inflation: Theory and observations}},  {\em Phys. Rept.} {\bf 402} (2004)
  103--266, [\href{http://xxx.lanl.gov/abs/astro-ph/0406398}{{\tt
  astro-ph/0406398}}].

\bibitem{2015arXiv150201592P}
{Planck Collaboration}, P.~A.~R. {Ade}, N.~{Aghanim}, M.~{Arnaud}, F.~{Arroja},
  M.~{Ashdown}, J.~{Aumont}, C.~{Baccigalupi}, M.~{Ballardini}, A.~J. {Banday},
  and et~al., {\it {Planck 2015 results. XVII. Constraints on primordial
  non-Gaussianity}},  {\em ArXiv e-prints} (Feb., 2015)
  [\href{http://xxx.lanl.gov/abs/1502.01592}{{\tt 1502.01592}}].

\bibitem{Dalal:2007cu}
N.~Dalal, O.~Dore, D.~Huterer, and A.~Shirokov, {\it {The imprints of
  primordial non-gaussianities on large-scale structure: scale dependent bias
  and abundance of virialized objects}},  {\em Phys. Rev.} {\bf D77} (2008)
  123514, [\href{http://xxx.lanl.gov/abs/0710.4560}{{\tt 0710.4560}}].

\bibitem{Slosar:2008hx}
A.~Slosar, C.~Hirata, U.~Seljak, S.~Ho, and N.~Padmanabhan, {\it {Constraints
  on local primordial non-Gaussianity from large scale structure}},  {\em JCAP}
  {\bf 0808} (2008) 031, [\href{http://xxx.lanl.gov/abs/0805.3580}{{\tt
  0805.3580}}].

\bibitem{Matarrese:2008nc}
S.~Matarrese and L.~Verde, {\it {The effect of primordial non-Gaussianity on
  halo bias}},  {\em Astrophys. J.} {\bf 677} (2008) L77--L80,
  [\href{http://xxx.lanl.gov/abs/0801.4826}{{\tt 0801.4826}}].

\bibitem{Wands:2010af}
D.~Wands, {\it {Local non-Gaussianity from inflation}},  {\em Class. Quant.
  Grav.} {\bf 27} (2010) 124002, [\href{http://xxx.lanl.gov/abs/1004.0818}{{\tt
  1004.0818}}].

\bibitem{Suyama:2010uj}
T.~Suyama, T.~Takahashi, M.~Yamaguchi, and S.~Yokoyama, {\it {On Classification
  of Models of Large Local-Type Non-Gaussianity}},  {\em JCAP} {\bf 1012}
  (2010) 030, [\href{http://xxx.lanl.gov/abs/1009.1979}{{\tt 1009.1979}}].

\bibitem{Byrnes:2010em}
C.~T. Byrnes and K.-Y. Choi, {\it {Review of local non-Gaussianity from
  multi-field inflation}},  {\em Adv. Astron.} {\bf 2010} (2010) 724525,
  [\href{http://xxx.lanl.gov/abs/1002.3110}{{\tt 1002.3110}}].

\bibitem{Koyama:2010xj}
K.~Koyama, {\it {Non-Gaussianity of quantum fields during inflation}},  {\em
  Class. Quant. Grav.} {\bf 27} (2010) 124001,
  [\href{http://xxx.lanl.gov/abs/1002.0600}{{\tt 1002.0600}}].

\bibitem{Chen:2010xka}
X.~Chen, {\it {Primordial Non-Gaussianities from Inflation Models}},  {\em Adv.
  Astron.} {\bf 2010} (2010) 638979,
  [\href{http://xxx.lanl.gov/abs/1002.1416}{{\tt 1002.1416}}].

\bibitem{Sefusatti:2007ih}
E.~Sefusatti and E.~Komatsu, {\it {The bispectrum of galaxies from
  high-redshift galaxy surveys: Primordial non-Gaussianity and non-linear
  galaxy bias}},  {\em Phys. Rev.} {\bf D76} (2007) 083004,
  [\href{http://xxx.lanl.gov/abs/0705.0343}{{\tt 0705.0343}}].

\bibitem{Sefusatti:2009qh}
E.~Sefusatti, {\it {1-loop Perturbative Corrections to the Matter and Galaxy
  Bispectrum with non-Gaussian Initial Conditions}},  {\em Phys. Rev.} {\bf
  D80} (2009) 123002, [\href{http://xxx.lanl.gov/abs/0905.0717}{{\tt
  0905.0717}}].

\bibitem{Yokoyama:2013mta}
S.~Yokoyama, T.~Matsubara, and A.~Taruya, {\it {Halo/galaxy bispectrum with
  primordial non-Gaussianity from integrated perturbation theory}},  {\em Phys.
  Rev.} {\bf D89} (2014), no.~4 043524,
  [\href{http://xxx.lanl.gov/abs/1310.4925}{{\tt 1310.4925}}].

\bibitem{Mizuno:2010by}
S.~Mizuno and K.~Koyama, {\it {Trispectrum estimator in equilateral type
  non-Gaussian models}},  {\em JCAP} {\bf 1010} (2010) 002,
  [\href{http://xxx.lanl.gov/abs/1007.1462}{{\tt 1007.1462}}].

\bibitem{Izumi:2011di}
K.~Izumi, S.~Mizuno, and K.~Koyama, {\it {Trispectrum estimation in various
  models of equilateral type non-Gaussianity}},  {\em Phys. Rev.} {\bf D85}
  (2012) 023521, [\href{http://xxx.lanl.gov/abs/1109.3746}{{\tt 1109.3746}}].

\bibitem{Senatore:2010jy}
L.~Senatore and M.~Zaldarriaga, {\it {A Naturally Large Four-Point Function in
  Single Field Inflation}},  {\em JCAP} {\bf 1101} (2011) 003,
  [\href{http://xxx.lanl.gov/abs/1004.1201}{{\tt 1004.1201}}].

\bibitem{Senatore:2010wk}
L.~Senatore and M.~Zaldarriaga, {\it {The Effective Field Theory of Multifield
  Inflation}},  {\em JHEP} {\bf 04} (2012) 024,
  [\href{http://xxx.lanl.gov/abs/1009.2093}{{\tt 1009.2093}}].

\bibitem{Smith:2015uia}
K.~M. Smith, L.~Senatore, and M.~Zaldarriaga, {\it {Optimal analysis of the CMB
  trispectrum}},  \href{http://xxx.lanl.gov/abs/1502.00635}{{\tt 1502.00635}}.

\bibitem{2015PhRvD..91l3521M}
S.~{Mizuno} and S.~{Yokoyama}, {\it {Halo/galaxy bispectrum with
  equilateral-type primordial trispectrum}},  {\em Phys. Rev.} {\bf 91} (June,
  2015) 123521, [\href{http://xxx.lanl.gov/abs/1504.05505}{{\tt 1504.05505}}].

\bibitem{Hashimoto:2015tnv}
I.~Hashimoto, A.~Taruya, T.~Matsubara, T.~Namikawa, and S.~Yokoyama, {\it
  {Constraining higher-order parameters for primordial non-Gaussianities from
  power spectra and bispectra of imaging surveys}},  {\em Phys. Rev.} {\bf D93}
  (2016), no.~10 103537, [\href{http://xxx.lanl.gov/abs/1512.08352}{{\tt
  1512.08352}}].

\bibitem{2015arXiv150201589P}
{Planck Collaboration}, P.~A.~R. {Ade}, N.~{Aghanim}, M.~{Arnaud},
  M.~{Ashdown}, J.~{Aumont}, C.~{Baccigalupi}, A.~J. {Banday}, R.~B.
  {Barreiro}, J.~G. {Bartlett}, and et~al., {\it {Planck 2015 results. XIII.
  Cosmological parameters}},  {\em ArXiv e-prints} (Feb., 2015)
  [\href{http://xxx.lanl.gov/abs/1502.01589}{{\tt 1502.01589}}].

\bibitem{Maldacena:2002vr}
J.~M. Maldacena, {\it {Non-Gaussian features of primordial fluctuations in
  single field inflationary models}},  {\em JHEP} {\bf 05} (2003) 013,
  [\href{http://xxx.lanl.gov/abs/astro-ph/0210603}{{\tt astro-ph/0210603}}].

\bibitem{Creminelli:2005hu}
P.~Creminelli, A.~Nicolis, L.~Senatore, M.~Tegmark, and M.~Zaldarriaga, {\it
  {Limits on non-gaussianities from wmap data}},  {\em JCAP} {\bf 0605} (2006)
  004, [\href{http://xxx.lanl.gov/abs/astro-ph/0509029}{{\tt
  astro-ph/0509029}}].

\bibitem{Huang:2006eha}
X.~Chen, M.-x. Huang, and G.~Shiu, {\it {The Inflationary Trispectrum for
  Models with Large Non-Gaussianities}},  {\em Phys. Rev.} {\bf D74} (2006)
  121301, [\href{http://xxx.lanl.gov/abs/hep-th/0610235}{{\tt
  hep-th/0610235}}].

\bibitem{Arroja:2009pd}
F.~Arroja, S.~Mizuno, K.~Koyama, and T.~Tanaka, {\it {On the full trispectrum
  in single field DBI-inflation}},  {\em Phys. Rev.} {\bf D80} (2009) 043527,
  [\href{http://xxx.lanl.gov/abs/0905.3641}{{\tt 0905.3641}}].

\bibitem{Chen:2009bc}
X.~Chen, B.~Hu, M.-x. Huang, G.~Shiu, and Y.~Wang, {\it {Large Primordial
  Trispectra in General Single Field Inflation}},  {\em JCAP} {\bf 0908} (2009)
  008, [\href{http://xxx.lanl.gov/abs/0905.3494}{{\tt 0905.3494}}].

\bibitem{2000ApJ...538..473L}
A.~{Lewis}, A.~{Challinor}, and A.~{Lasenby}, {\it {Efficient Computation of
  Cosmic Microwave Background Anisotropies in Closed Friedmann-Robertson-Walker
  Models}},  {\em Astrophys. J.} {\bf 538} (Aug., 2000) 473--476,
  [\href{http://xxx.lanl.gov/abs/astro-ph/9911177}{{\tt astro-ph/9911177}}].

\bibitem{2011PhRvD..83h3518M}
T.~{Matsubara}, {\it {Nonlinear perturbation theory integrated with nonlocal
  bias, redshift-space distortions, and primordial non-Gaussianity}},  {\em
  Phys. Rev.} {\bf 83} (Apr., 2011) 083518,
  [\href{http://xxx.lanl.gov/abs/1102.4619}{{\tt 1102.4619}}].

\bibitem{2012PhRvD..86f3518M}
T.~{Matsubara}, {\it {Deriving an accurate formula of scale-dependent bias with
  primordial non-Gaussianity: An application of the integrated perturbation
  theory}},  {\em Phys. Rev.} {\bf 86} (Sept., 2012) 063518,
  [\href{http://xxx.lanl.gov/abs/1206.0562}{{\tt 1206.0562}}].

\bibitem{2014PhRvD..89d3524Y}
S.~{Yokoyama}, T.~{Matsubara}, and A.~{Taruya}, {\it {Halo/galaxy bispectrum
  with primordial non-Gaussianity from integrated perturbation theory}},  {\em
  Phys. Rev.} {\bf 89} (Feb., 2014) 043524,
  [\href{http://xxx.lanl.gov/abs/1310.4925}{{\tt 1310.4925}}].

\bibitem{2001MNRAS.323....1S}
R.~K. {Sheth}, H.~J. {Mo}, and G.~{Tormen}, {\it {Ellipsoidal collapse and an
  improved model for the number and spatial distribution of dark matter
  haloes}},  {\em Mon. Not. Roy. Astron. Soc.} {\bf 323} (May, 2001) 1--12,
  [\href{http://xxx.lanl.gov/abs/astro-ph/9907024}{{\tt astro-ph/9907024}}].

\bibitem{2011PhRvD..83l3514N}
T.~{Namikawa}, T.~{Okamura}, and A.~{Taruya}, {\it {Magnification effect on the
  detection of primordial non-Gaussianity from photometric surveys}},  {\em
  Phys. Rev.} {\bf 83} (June, 2011) 123514,
  [\href{http://xxx.lanl.gov/abs/1103.1118}{{\tt 1103.1118}}].

\bibitem{2008PhRvD..78l3506L}
M.~{Loverde} and N.~{Afshordi}, {\it {Extended Limber approximation}},  {\em
  Phys. Rev.} {\bf 78} (Dec., 2008) 123506,
  [\href{http://xxx.lanl.gov/abs/0809.5112}{{\tt 0809.5112}}].

\bibitem{1954ApJ...119..655L}
D.~N. {Limber}, {\it {The Analysis of Counts of the Extragalactic Nebulae in
  Terms of a Fluctuating Density Field. II.}},  {\em Astrophys. J.} {\bf 119}
  (May, 1954) 655.

\bibitem{HSCrev}
H.~Collaboration, {\it {Hyper Suprime-Cam Design Review}}, .
  \verb|http://www.naoj.org/Projects/HSC/j_report.html|.

\bibitem{2005astro.ph.10346T}
{The Dark Energy Survey Collaboration}, {\it {The Dark Energy Survey}},  {\em
  ArXiv Astrophysics e-prints} (Oct., 2005)
  [\href{http://xxx.lanl.gov/abs/astro-ph/0510346}{{\tt astro-ph/0510346}}].

\bibitem{2009arXiv0912.0201L}
{LSST Science Collaboration}, P.~A. {Abell}, J.~{Allison}, S.~F. {Anderson},
  J.~R. {Andrew}, J.~R.~P. {Angel}, L.~{Armus}, D.~{Arnett}, S.~J. {Asztalos},
  T.~S. {Axelrod}, and et~al., {\it {LSST Science Book, Version 2.0}},  {\em
  ArXiv e-prints} (Dec., 2009) [\href{http://xxx.lanl.gov/abs/0912.0201}{{\tt
  0912.0201}}].

\bibitem{2013MNRAS.429..344K}
I.~{Kayo}, M.~{Takada}, and B.~{Jain}, {\it {Information content of weak
  lensing power spectrum and bispectrum: including the non-Gaussian error
  covariance matrix}},  {\em Mon. Not. Roy. Astron. Soc.} {\bf 429} (Feb.,
  2013) 344--371, [\href{http://xxx.lanl.gov/abs/1207.6322}{{\tt 1207.6322}}].

\bibitem{2013PhR...530...87W}
D.~H. {Weinberg}, M.~J. {Mortonson}, D.~J. {Eisenstein}, C.~{Hirata}, A.~G.
  {Riess}, and E.~{Rozo}, {\it {Observational probes of cosmic acceleration}},
  {\em Phys.rep} {\bf 530} (Sept., 2013) 87--255,
  [\href{http://xxx.lanl.gov/abs/1201.2434}{{\tt 1201.2434}}].

\bibitem{2013MNRAS.432.2945H}
D.~{Huterer}, C.~E. {Cunha}, and W.~{Fang}, {\it {Calibration errors unleashed:
  effects on cosmological parameters and requirements for large-scale structure
  surveys}},  {\em Mon. Not. Roy. Astron. Soc.} {\bf 432} (July, 2013)
  2945--2961, [\href{http://xxx.lanl.gov/abs/1211.1015}{{\tt 1211.1015}}].

\end{thebibliography}\endgroup

\end{document}